%% file: main.tex
\documentclass[letterpaper,twocolumn,10pt]{article}
\usepackage{usenix-2020-09}
\usepackage{tikz}
\usepackage{amsthm}
\usepackage{amssymb}
\usepackage{amsfonts}

\usepackage{cite}

\usepackage{blindtext}
\usepackage{xcolor}

\usepackage{graphicx}
\usepackage{enumitem}
\usepackage{adjustbox}
\setlist[enumerate]{leftmargin=*}
\setlist[itemize]{leftmargin=*}
\setlist{nolistsep}
\usepackage{amsthm}

\newtheorem{definition}{Definition}

\usepackage{framed}
\definecolor{shadecolor}{rgb}{0,1,0}

\usepackage{pifont}
\usepackage{verbatim}
\usepackage{subfigure} 
\usepackage{threeparttable}
\usepackage{tikz}
\newcommand*\emptycirc[1][0.8ex]{\tikz\draw (0,0) circle (#1);} 
\newcommand*\halfcirc[1][0.8ex]{%
  \begin{tikzpicture}
  \draw[fill] (0,0)-- (90:#1) arc (90:270:#1) -- cycle ;
  \draw (0,0) circle (#1);
  \end{tikzpicture}}
\newcommand*\fullcirc[1][0.8ex]{\tikz\fill (0,0) circle (#1);}

\usepackage[cmex10]{amsmath}
\usepackage{mdframed}

\usepackage{algorithm}
\usepackage{booktabs}

\usepackage{algorithm}
\usepackage{algorithmic}

\usepackage{bbding}

\usepackage{makecell}
\usepackage{multirow}
\usepackage{enumitem}

\usepackage{lipsum}

\begin{document}
%
\title{Rethinking White-Box Watermarks on Deep Learning Models \\ under Neural Structural Obfuscation}



\author{\rm{Yifan Yan*, Xudong Pan*, Mi Zhang\textsuperscript{\Envelope}, Min Yang\textsuperscript{\Envelope}} \\ \textit{Fudan University, China} \\
\{yanyf20, xdpan18, mi\_zhang, m\_yang\}@fudan.edu.cn  \\ {\small{(*: co-first authors; \Envelope: corresponding authors)}} 
} 
\maketitle

\input{tex/abstract.tex}

\input{tex/intro.tex}
\input{tex/related.tex}
\input{tex/prelim.tex}
\input{tex/threat_model.tex}
\input{tex/motive.tex}
\input{tex/attack.tex}

\input{tex/evaluation.tex}

\input{tex/discussion.tex}

\input{tex/cls.tex}
\input{tex/ack.tex}
\bibliographystyle{unsrt}
\bibliography{ref}
%

\input{tex/appendix.tex}

\end{document}

%% file: tex/abstract.tex
\begin{abstract}
          
Copyright protection for deep neural networks (DNNs) is an urgent need for AI corporations. To trace illegally distributed model copies, DNN watermarking is an emerging technique for embedding and verifying secret identity messages in the prediction behaviors or the model internals. Sacrificing less functionality and involving more knowledge about the target DNN, the latter branch called \textit{white-box DNN watermarking} is believed to be accurate, credible and secure against most known watermark removal attacks, with emerging research efforts in both the academy and the industry.

In this paper, we present the first systematic study on how the mainstream white-box DNN watermarks are commonly vulnerable to neural structural obfuscation with \textit{dummy neurons}, a group of neurons which can be added to a target model but leave the model behavior invariant. Devising a comprehensive framework to automatically generate and inject dummy neurons with high stealthiness, our novel attack intensively modifies the architecture of the target model to inhibit the success of watermark verification. With extensive evaluation, our work for the first time shows that nine published watermarking schemes require amendments to their verification procedures.

\end{abstract}

%% file: tex/intro.tex
\section{Introduction}
Nowadays, the computational and engineering costs of training a giant DNN model increase faster than ever \cite{he2016resnet, xu2015caption,devlin2018bert, strubell2019energy}. As a critical asset of AI corporations, well-trained DNNs are exposed under the risk of model stealing attacks \cite{oh2019reverseNN, orekondy2019knockoff, wang2018stealhyper, yan2020cache,Jagielski2020HighAA,Zhu2021HermesAS}, which makes the need for model copyright protection current and pressing. As a rescue, the past few years witness the emergence of DNN watermarking \cite{uchida2017embedding, wang2021riga,liu2021greedyresiduals, fan2021deepip, zhang2020passportaware, ong2021iprgan,chen2021lottery, lim2022ipcaption,darvish2019deepsigns,adi2018turning, szyller2021dawn, jia2021entangled} for tracing illegal model copies in the wild \cite{sokwatermark}. 
Generally, a model watermarking scheme consists of \textit{watermark embedding} and \textit{verification}. At the former stage, a secret identity message, i.e., the \textit{watermark}, is first embedded into \textit{the target model} along with the training process. At the latter stage, the scheme verifies the ownership according to whether the same or a similar watermark is detected from \textit{a suspect model}. 

According to the location of the embedded message, existing DNN watermarks are categorized into \textit{black-box} and \textit{white-box}. Intuitively, a black-box watermark is embedded in the model's prediction behavior on a special set of inputs \cite {adi2018turning,zhang2018blackwatermark, szyller2021dawn, jia2021entangled}, while a white-box watermark is embedded in the model internals, including the model parameters \cite{uchida2017embedding,  wang2021riga,liu2021greedyresiduals, chen2021lottery,fan2021deepip, zhang2020passportaware, ong2021iprgan} and the neuron activation \cite{darvish2019deepsigns, lim2022ipcaption}. The difference above also determines the required access mode to the suspect model for verification. As suggested by Fan et al. \cite{fan2021deepip}, in a real-world copyright dispute, the owner may first collect evidence of model piracy via a black-box query and then attain the white-box access via law enforcement for ownership verification.

\input{tex/tables/intro_table_ccs.tex}


Sacrificing less functionality and involving more information for verification, white-box model watermarks are widely considered more comprehensive compared with the black-box counterpart \cite{wang2021riga,liu2021greedyresiduals, fan2021deepip, jia2021entangled}, with increasingly more research efforts on top-tier AI/security/system venues and from industry leaders (e.g., Microsoft \cite{zhang2020passportaware, darvish2019deepsigns, chen2019deepattest}). In a typical attack scenario, the adversary with a stolen DNN would modify the parameters or the structure of the model to frustrate the success of watermark verification \cite{see2016compression, li2016filterpruning, hinton2015distilling, wang2019overwrite, yang2019distillremoval, shafieinejad2021robustofbackdoorbased, chen2021refit, aiken2021laundering, guo2021ftnotenough, wang2019neuralcleanse}. To achieve the attack goal, the primary constraints for the attacker are (i) the obfuscation process should not cost more \textit{resources} than training a DNN from scratch and (ii) the \textit{utility} of the obfuscated model should have no clear decrease.

However, as summarized in Table \ref{tab:intro_table_comparison}, none of the existing approaches can balance well the cost on utility or computing resources for fully removing the embedded watermark. On the one hand, removal attacks by parameter modification  
inevitably encounter degradation in the normal model utility \cite{adi2018turning,zhang2020passportaware,fan2021deepip,liu2021greedyresiduals,ong2021iprgan,chen2021lottery}. Relying on the internals of the suspect model, the embedded identity messages in white-box watermarking are much strongly connected with the model performance. 
Therefore, attack attempts via conventional post-processing techniques  \cite{see2016compression, li2016filterpruning}, which show empirical success on black-box model watermarks, inevitably perturb the model parameters at an unacceptable scale to fully remove a white-box model watermark \cite{uchida2017embedding, wang2021riga,liu2021greedyresiduals, chen2021lottery, fan2021deepip, zhang2020passportaware, ong2021iprgan,darvish2019deepsigns, lim2022ipcaption}. On the other hand, existing structural modification attacks apply knowledge distillation on the target model to construct a substitute model with similar performance but of different neural architecture   \cite{hinton2015distilling,Gou2021KnowledgeDA}. However, they usually require additional computational resources for training the substitute model. Besides, some attacks further require the access to a domain dataset or require additional knowledge about the embedded watermark \cite{wang2021riga}, which are usually impractical for attacks in the wild.  

\noindent\textbf{Our Work.} We for the first time show, most of the state-of-the-art white-box DNN watermarks share common vulnerabilities in their verification procedures which assume the structural integrity of the suspect model after being obfuscated by the attacker. Our current work constructs a novel neural structural obfuscation attack which intensively modifies the architecture of the victim model to disable the verification procedures of nine previously published schemes. Meanwhile, our attack incurs no utility loss and training costs, and requires neither dataset access nor the knowledge about the embedded watermark. 
At the core of our newly proposed attack is the concept of \textit{dummy neurons}, literally a group of neurons which can be added to a target DNN model for intensively perturbing the embedded watermark while provably leaving the model behavior \textit{invariant} (i.e., the model output remains the same under the same input). A naive example is neurons which have the input and output weights of zero values, which, if added to a DNN model, have no contribution to its output. As a preliminary yet effective attack, the adversary obfuscates the protected model by injecting a number of these neurons to every neural layer, which already inhibits most of the state-of-the-art white-box watermarks from being executed, but has clear limitation in its attack stealthiness (\S\ref{sec:motivation}).





Alternatively, we propose a more comprehensive attack framework to automatically generate and inject dummy neurons into a victim model, which implements by-design stealthiness of the injected dummy neurons when the obfuscated model is under inspection. For dummy neuron generation, we propose \textit{NeuronClique} and \textit{NeuronSplit}, two novel structural obfuscation primitives to construct groups of dummy neurons, where the neurons are associated with non-vanishing weights but still bring no change to the model output. Specifically, the \textit{NeuronClique} primitive directly generates an arbitrary number of neurons which are assigned with weights that can cancel the others' output out, while \textit{NeuronSplit} converts a neuron in the victim model into two substitute neurons which preserve the replaced neuron's functionality (\S\ref{sec:dn_generation}).

For dummy neuron injection, our proposed framework carefully designs the injection order and leverages the reciprocity between dummy neurons in successive layers to enhance the attack stealthiness (\S\ref{sec:dn_injection}). Furthermore, we leverage the scaling invariance in DNN \cite{neyshabur2015pathsgd} to provide the adversary with the flexibility to specify the weight distribution of the dummy neurons to follow the same distribution of the original neurons, and the shuffling invariance in DNN \cite{ganju2018property} to randomize the location of the injected dummy neurons among the original neurons. Finally, we also introduce the kernel expansion technique to further obfuscate the weight shape of the dummy neurons, which, as the final straw, turns the victim model into a structurally irrelevant model with its original self (\S\ref{sec:dn_camouflage}). In \S\ref{sec:eval:stealthiness_dn}, we discuss and experiment with the feasibility for a defender of different knowledge on our attack to attempt to remove the dummy neurons.

\noindent\textbf{Our Contributions.} In summary, we mainly make the following contributions:
\begin{itemize}
\item We for the first time reveal the common vulnerability of the state-of-the-art white-box DNN watermarks to neural structural obfuscation with dummy neurons.

\item We devise a comprehensive attack framework which automatically generates groups of dummy neurons into a protected model with newly proposed attack primitives. 

 
\item We validate the success of our attack on a wide group of DNNs protected by nine published white-box watermarking schemes. Despite the claimed robustness, the success rate of watermark verification is reduced to random after our attack, while the normal model utility remains the same.

\item We also provide a study on the stealthiness of these dummy neurons and present a dummy neuron elimination algorithm. This possible defense eliminates the dummy neurons, while the original model watermark in the protected model is recovered only when the defender has access to the original watermarked model.

\end{itemize}


%% file: tex/tables/intro_table_ccs.tex
\begin{table}[t]
  \caption{Compared with existing attacks, our attack is the first to disable the verification procedures of nine state-of-the-art white-box watermarks under no requirements on utility loss, dataset access, training costs or watermark knowledge.}
    \centering
  \scalebox{0.7}{
      \begin{tabular}{llllll}
    \toprule
         \multicolumn{1}{p{4.19em}}{\textbf{Attack\newline{}Type}}  & \multicolumn{1}{p{4.19em}}{\textbf{Attack\newline{}Class}} & \multicolumn{1}{p{4.19em}}{\textbf{Utility\newline{}Loss}} & \multicolumn{1}{p{4.19em}}{\textbf{Training\newline{}Cost}} & \multicolumn{1}{p{4.19em}}{\textbf{Dataset\newline{}Access}} & \multicolumn{1}{p{4.19em}}{\textbf{Watermark\newline{}Knowledge}} \\
    \midrule
    \textbf{Pruning} & Parameter  &    \fullcirc   &  \emptycirc       &  \emptycirc &  \halfcirc \\
    \textbf{Finetuning} & Parameter    &      \emptycirc  &  \fullcirc      &  \fullcirc &  \halfcirc \\
    \textbf{Overwriting} & Parameter      &  \halfcirc      &  \halfcirc      &  \fullcirc &  \fullcirc \\
    \textbf{Extraction} & Structure      &   \halfcirc     & \halfcirc       & \fullcirc &  \emptycirc \\
    \textbf{Ours} & Structure  &  \emptycirc      &    \emptycirc    & \emptycirc  &  \emptycirc \\
    \bottomrule
    \end{tabular}}%
    
     \begin{tablenotes}
        \footnotesize
 \item *\fullcirc/\halfcirc/\emptycirc\text{ }denote large/moderate/no tradeoff in each dimension.
      \end{tablenotes}
  \label{tab:intro_table_comparison}%
\vspace{-0.3in}
\end{table}%


%% file: tex/related.tex
\section{Related Work}
\noindent\textbf{DNN Watermarking Schemes.}
Two categories of DNN watermarking methods, i.e., black-box and white-box algorithms, have been proposed to support model ownership verification. The black-box watermark schemes \cite{adi2018turning, jia2021entangled} mostly embed the identity information into the input-output patterns of the target model on a secret trigger set (similar to backdoor attacks \cite{gu2017badnets}).  As reported in \cite{adi2018turning}, the trade-off is sometimes evident between successfully embedding a black-box watermark and preserving the correct predictions on normal inputs. Moreover, recent progress on backdoor defenses also exposes a new attack surface on these black-box watermark schemes \cite{liu2019abs, liu2018finepruning, wang2019neuralcleanse}. 
White-box watermarking requires access to the parameters or the neuron activation of the protected model to extract the watermark. According to the location of the embedded message, the white-box watermarking methods can be classified into three groups: weight-based \cite{uchida2017embedding, wang2021riga, ong2021iprgan, liu2021greedyresiduals, chen2021lottery}, activation-based\cite{darvish2019deepsigns, lim2022ipcaption}, and passport-based\cite{fan2021deepip, zhang2020passportaware}. Recent watermarking schemes always show strong robustness against existing removal attacks including fine-tuning, pruning and overwriting \cite{wang2019overwrite}. Very recently, a concurrent work by Yan et al. \cite{yan2022cracking} reveals the overly dependence of existing white-box watermarks on the local neuron features which are fragile under neuron permutation and rescaling, while our revealed vulnerability is rooted in their common dependence on the structural identity of the target and the suspect model.

As the black-box and white-box watermarking schemes do not conflict with each other, some recent works combine them to provide more robust protection to the model copyright \cite{chen2021lottery,fan2021deepip,zhang2020passportaware,ong2021iprgan,darvish2019deepsigns}. During their watermark verification, these hybrid watermark algorithms first collect sufficient evidence via remote queries to the suspect models. Then, the owner further attains full access to the model with law enforcement to detect the identity information in the model internals, which yields a strong copyright statement. 

\noindent\textbf{Program Obfuscation.}
To prevent data structures and control flow of source code from being exposed through reverse engineering attacks, program obfuscation transforms a computer program that is semantic-equivalent to the original one but is harder to be analyzed for protecting the confidentiality of the program internals \cite{cifuentes1995decompilation, tip1994surveyanalyze}. This prevents an attacker or an analyst from reverse-engineering or debugging a proprietary software program \cite{xu2017ProgramObfuscationSurvey, schrittwieser2016canitobfuscation} via layout transformation, control-flow transformation, or data obfuscations \cite{collberg1997taxonomyobfuscating,majumdar2006surveyobfuscation,drape2010intellectualobfuscation}. 

Our proposed neural structural obfuscation is designed for a similar goal as program obfuscation, i.e., to prevent the copyright verification algorithm from successfully validating the watermark existence. Technically different from program obfuscation, neural structural obfuscation is done via different structurally invariant transforms on a neural network protected by the white-box watermarking. The obfuscated neural network is functionally equivalent to the original one, while the existing verification procedures can no longer recognize the original watermark from the model.

%% file: tex/prelim.tex
\section{Preliminary}

\noindent\textbf{Notations on Deep Neural Networks.} Considering a DNN with $H$ layers, i.e., $\{f^1, f^2, \hdots,f^H\}$, each layer $f^l$ is composed of a set of $N_l$ neurons ($f^l=(n_1^l, n_2^l,...,n_{N_l}^l)$). We denote the parameters of the $l^{th}$ $(1\leq l\leq H)$ layer as $W^l$,
which can be further written as $\{w^l_{ij}\}_{i=1,j=1}^{N_{l-1}, N_l} \cup \{b^l_{j}\}_{j=1}^{N_l}$. From the neuron-level viewpoint, each element $w^l_{ij}$ is a scalar value (i.e., \textit{weight}) in a linear layer, or a matrix (i.e., \textit{kernel}) in a convolutional layer, which connects the neurons $n^{l-1}_i$ and $n^l_j$, and $b^l_{j}$ is the bias.



\noindent\textbf{White-box DNN Watermarking.}
As an effective forensic technique against model stealing attacks \cite{oh2019reverseNN, orekondy2019knockoff, wang2018stealhyper, yan2020cache,Jagielski2020HighAA,Zhu2021HermesAS}, a number of model watermark schemes are proposed from 2017 \cite{uchida2017embedding, wang2021riga,liu2021greedyresiduals, fan2021deepip, zhang2020passportaware, ong2021iprgan,chen2021lottery, lim2022ipcaption,darvish2019deepsigns,adi2018turning,zhang2018blackwatermark, szyller2021dawn, jia2021entangled}, which allow the legitimate model owner to establish the legal ownership by verifying the existence of a unique watermark, usually in the form of secret identity messages, in the suspect model. Our current work concentrates on the security analysis of white-box watermarking schemes, an important and evolving branch of DNN watermarking which embeds and verifies the watermark in model internals (i.e., parameters \cite{uchida2017embedding,  wang2021riga,liu2021greedyresiduals, chen2021lottery,fan2021deepip, zhang2020passportaware, ong2021iprgan} or activation maps \cite{darvish2019deepsigns, lim2022ipcaption}).   

Broadly speaking, a white-box model watermark scheme consists of the following phases: \textit{watermark embedding} and \textit{watermark verification}. In the former phase, the scheme embeds the secret message $s$ (e.g., a bit string) into the parameters or the intermediate activation maps of the owned model $f_W$. This is usually achieved by an additional regularization term $\mathcal{L}_{\text{wmk}}$ alongside the primary learning objective $\mathcal{L}$, i.e.,
$
\mathcal{L}' = \mathcal{L} + \lambda \mathcal{L}_{\text{wmk}},
$
where $\lambda$ is the hyper-parameter to balance the utility and the specificity of the embedded watermark. For example, Uchida et al. \cite{uchida2017embedding} embed a secret bit string $s$ in the kernels of a specified convolutional layer. Therefore, the regularization term $\mathcal{L}_{wmk}$ is designed as the binary cross entropy loss between the secret bit string $s$ and $\sigma(X\cdot w)$, where $w$ is derived from the parameters of a specified convolutional layer by channel-level averaging and flattening, $X$ is a predefined transformation matrix, and $\sigma$ is the sigmoid function. During the verification, a watermark extraction function $\mathcal{E}$ is implemented to extract an equal-length message $s'$ from a given suspect model  $f_{\tilde{W}}$ in polynomial time: $\label{eq:extract wm}
s' = \mathcal{E}(f_{\tilde{W}}, M, A),
$
where $M$ is the mask matrix to select a set of the specific parameters or activation maps directly from the suspect model $f_{\tilde{W}}$, and $A$ is a transformation function which projects the selected weights or activation maps to obtain the extracted message. 




 To ensure a trustworthy ownership verification, a model watermark scheme should satisfy the minimum set of requirements. For more advanced model watermark requirements, please refer to \cite{sokwatermark}.
\begin{itemize}
\item \textbf{Fidelity:} The utility of the model should have as small decrease as possible when the watermark is embedded.

\item \textbf{Reliability:} The watermark should be verified with high confidence in \textit{positive suspect models}, i.e., the same or a post-processed copy of the watermarked model, and with low confidence in negative suspect models, i.e., irrelevant models owned by others.

\item \textbf{Robustness:} The embedded watermark in the protected model should be resistant to adversarial attempts which aim at removing the watermark from the model (\S\ref{sec:limitations}). Moreover, a watermarking scheme should also raise the bar for the adversary to embed another fabricated watermark into the target model to cause ownership ambiguity \cite{fan2021deepip}.
\end{itemize}

%% file: tex/threat_model.tex
\section{Security of White-Box DNN Watermark} 
\subsection{Security Settings}
\noindent\textbf{Attack Taxonomy.} According to the adversarial goal, we first categorize existing attacks on white-box model watermarking from previous works into \textit{ambiguity attacks} and \textit{removal attacks}. In the former attacks, the adversary aims at constructing a counterfeit watermark, when given the watermarked DNN \cite{zhang2020passportaware}, to pass the verification process. Instead, the removal attacks have a more straightforward goal: invalidating the verification process by removing the secret identity message from the protected model. Considering its severe influence on establishing the model ownership, our work concentrates on devising novel removal attacks to crack the state-of-the-art white-box model watermarks. Below, we formally describe the attack scenario. 

\noindent\textbf{Attack Scenario.} In our threat model, the adversary has obtained an illegal copy of a watermarked model which allows full access to its model parameters. Such model piracy can be accomplished via either algorithmic attacks \cite{tramer2016StealViaApi, yu2020cloudleak} or system attacks exploiting software/hardware vulnerabilities \cite{jeong2021meltdown,yan2020cache}. To conceal the traces of model infringement, the attacker attempts to invalidate the model ownership verification by removing the existing watermarks. 

\noindent\textbf{Attack Budget.} As listed in Table \ref{tab:intro_table_comparison}, the attack budget of a removal attack is mainly measured in the following dimensions: \textit{utility loss}, \textit{training cost}, \textit{dataset access}, \textit{watermark knowledge} (similar to the ones on black-box watermark in \cite{sokwatermark}).

\begin{itemize}
\item \textbf{Utility Loss:} When removing the watermark, obfuscation on the parameters or the structure of the DNN model seems inevitable. In this case, the obfuscation should not incur a large decrease in the normal model utility, which is otherwise unacceptable because the attacker still requires the normal utility of the pirated model for profits.
\item \textbf{Training Cost:} For watermark removal, the adversary is usually unwilling to cost a similar scale of computing resources as retraining the DNN model from scratch. Typically, the adversary would avoid the expensive model training process, which usually involves the usage of high-end graphical cards for training industry-level models, but prefer to learning-free attacks.  

\item \textbf{Dataset Access:} As the training is usually a private asset of the model owner, the access to the original training data or even a public domain dataset brings an additional attack budget. The adversary would prefer to involve no real data inputs for conducting the attack.  

\item \textbf{Watermark Knowledge:} The adversary should have no knowledge about the adopted watermark embedding and extraction algorithms, which are usually exclusively known to the owner until the ownership verification is launched. 

\end{itemize}

\subsection{Limitation of Existing Removal Attacks}
\label{sec:limitations}
Previous removal attacks
are all limited in one or more of the above dimensions for fully removing white-box watermarks from a protected model. 

\begin{itemize}

\item \textbf{Pruning} sets a proportion of redundant parameters in DNN to zero, under which previous white-box watermark is highly resistant. To fully remove the watermark, pruning has to remove a substantial amount of weights, which causes an unacceptable utility loss \cite{uchida2017embedding, darvish2019deepsigns}.

\item \textbf{Finetuning} continues the training operation for a few epochs without the watermark-related loss. This removal attack additionally requires a certain amount of domain data and computational resources, otherwise the model utility would degrade \cite{chen2021refit, guo2021ftnotenough}.

\item \textbf{Overwriting} is first proposed in \cite{wang2019overwrite} to show the vulnerability of  \cite{uchida2017embedding}. Considering the adversary has full knowledge about the watermarking process, he/she may confuse the verification by embedding his own identification information. However, the details of watermark schemes are always not available in real-world settings. Meanwhile, overwriting attacks usually fail to encode a new message into the target model following more advanced schemes \cite{fan2021deepip, ong2021iprgan, lim2022ipcaption, zhang2020passportaware}.

\item \textbf{Extraction} refers to an attack class which utilizes knowledge distillation techniques \cite{hinton2015distilling} on the pirated model to obtain an obfuscated model which usually has a different architecture. The extraction attack inevitably involves a substantial amount of training costs to distill a well-trained obfuscated model. Although some very recent works in knowledge distillation eliminate the assumption on dataset access \cite{Yin2020DreamingTD}, most mainstream extraction attacks still use the conventional knowledge distillation approaches and require the access to a domain dataset to reduce the utility loss.     
\end{itemize}

%% file: tex/motive.tex
\section{A Motivating Example}
\label{sec:motivation}
\noindent\textbf{What is \textit{Dummy Neurons}?} Literally, dummy neurons are defined to be those neurons which leave the prediction behavior of the original DNN intact after being inserted into the model. 

\begin{definition}[Dummy Neurons]
When we insert a group of neurons $\{m_i\}^{M}_{i=1}$ into a given DNN $f$ to obtain a structurally obfuscated model $f^{'}$, we call the group of neurons \textit{dummy neurons} if for every input ${x} \in \mathcal{X}$, $f(x) = f^{'}(x)$.

\end{definition}


\begin{figure}[t]
\begin{center}
\includegraphics[width=0.45\textwidth]{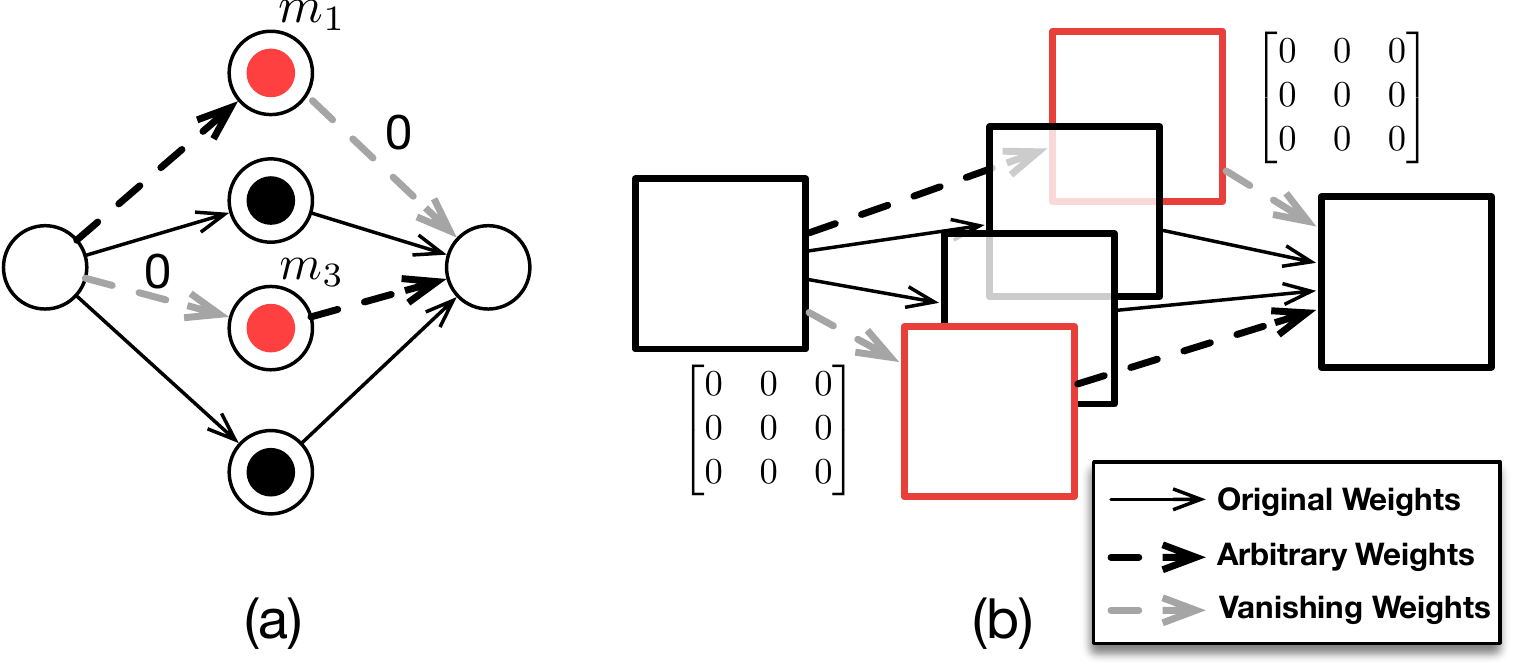}
\caption{A schematic diagram of NeuronZero on (a) fully-connected layers and (b) convolutional layers.}
\label{fig:naive_dn}
\end{center}
\end{figure}

\begin{figure*}[t]
\begin{center}
\includegraphics[width=0.9\textwidth]{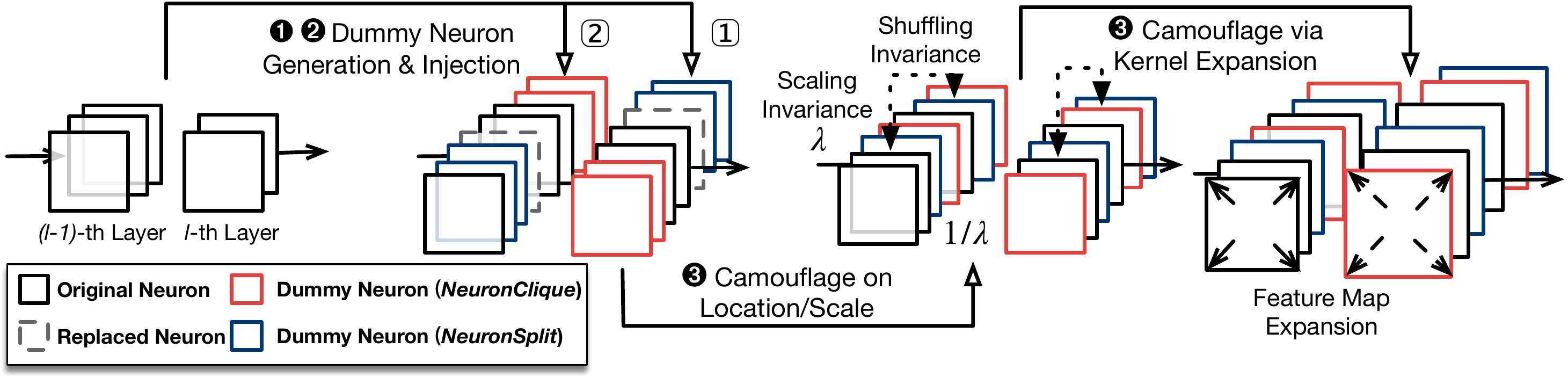}
\caption{Overview of our proposed watermark removal attack by neural structural obfuscation with dummy neurons.}
\label{fig:atk_pipeline}
\end{center}
\end{figure*}

We present motivating examples of dummy neurons in Fig.\ref{fig:naive_dn}. In Fig.\ref{fig:naive_dn}(a), given the target fully-connected neural network (FCN), we inject two additional neurons (i.e., $m_1$ and $m_3$) into each hidden layer of the model. By definition, each injected neuron $m_k$ is associated with vectors of incoming and outgoing weights, which connects the $k$-th dummy neuron with the $i$-th neuron in the precedent layer and $j$-th neuron in the successive layer (denoted as $u_{i,k}$ and $v_{k,j}$, respectively). According to the architecture specification, a neuron may optionally be associated with a bias $o^l_k$. 

To make the injected neurons dummy, a naive strategy is to set the incoming/outgoing weight of the neuron to be a vector of all $0$, the optional bias $0$ but leave the incoming/outgoing weights arbitrarily assigned. For example, the dummy neuron $m_1$ in Fig.\ref{fig:naive_dn}(a) has the outgoing weights of values $0$ (i.e., $v_{1,j} = 0$ for $j$). Therefore, its contribution of $m_1$ to any of the neurons in the successive layer is constantly $0$. Alternatively, the dummy neuron $m_3$ has the incoming weights of values $0$, which implies that the activation of $m_3$ is constantly $0$. Both cases provably ensure no impact on the next layer's output and finally leaves the utility of the model intact. As a fully connected layer in DNNs is a simplified form of convolutional layer, we can add the dummy neurons in convolutional neural network (CNN) in the same way (Fig.\ref{fig:naive_dn}(b)), i.e., setting the weights of all the incoming or outgoing filters to $0$. We refer to the above construction of the dummy neurons as \textit{NeuronZero}.

\noindent$\bullet$\textbf{ A Preliminary Attack.} With \textit{NeuronZero}, the adversary is ready to obfuscate the protected model by injecting an arbitrary number of dummy neurons with incoming/outgoing weights of values $0$ to every neural layer. The process requires no knowledge about whether and what type of watermark is embedded, and does not need to train the protected model. As we analyze below, the attack is also effective to invalidate the mainstream white-box model watermark schemes by perturbing the watermark-related parameters.

\noindent\textbf{Procedural Vulnerability of White-Box Watermark} After dummy neurons are injected into each neural layer of the protected model, the parameters once embedded with the watermark information are now messed with the weights of the dummy neurons. Consequently, when the verification process invokes the watermark extraction algorithm on the tampered weights, the extracted watermark information is very likely to suffer a large distortion. Worse yet, most of the state-of-art white-box watermarks do not implement error-handling procedures to properly address the case when the size of the weights from the watermarked layer mismatches the expected size in the watermark extraction algorithm. Therefore, these watermark schemes are inexecutable for extracting valid watermarks from the suspect model and thus can no longer protect the intellectual property of the victim model. 

\noindent$\bullet$\textbf{ An Example.} Uchida et al. embed the watermark into the convolutional filter weights of a target layer, i.e., $W \in \mathbb{R}^{n\times c \times w \times h}$, where $n$ is the number of filters, $c$ is the number of channels, and $w$, $h$ are the width and height of these filters\cite{uchida2017embedding}. To extract the watermark, the verification process first averages these convolutional weights at channel level and then flattens the result to $\hat W \in \mathbb{R}^{c \cdot w \cdot h}$. Finally, a bit string with length of $b$ is obtained according to the signs of $\hat W$, i.e., $s' = T_h(A\cdot \hat W) = \{0,1\}^b$, where $T_h$ is a hard threshold function which output $1$ when the input is positive and $0$ otherwise, and $A\in \mathbb{R}^{b \times (c \cdot w \cdot h)}$ is a pre-defined transform matrix sampled from the normal distribution. Once we add a group of dummy neurons into the next layer of watermarked layer (e.g., Fig.\ref{fig:naive_dn}(b)), the weight $W'\in \mathbb{R}^{n' \times c' \times w \times h}$ in the watermarked layer now has an expanded shape, i.e., the numbers of filters and channels in the target layer increase to $n', c'$ due to the injected dummy neurons. After being flattened, the parameters become $\hat W' \in \mathbb{R}^{c' \cdot w \cdot h}$ in vector form. As a result, the verification process cannot be executed as the dimension of $\hat W'$ is incompatible with the second dimension of the transform matrix $A\in \mathbb{R}^{b \times (c \cdot w \cdot h)}$. By error-handling the flattened weight $\hat W'$ via, e.g., random sampling or max pooling, to satisfy the dimension consistency with $A$, a bit string can still be obtained from the watermarked layer. However, as validated in Section \ref{sec:eval_weight}, the extracted message remains almost at random unless the verifier were able to precisely eliminate the impact of the dummy neurons from $\hat W'$ to restore the watermark integrity.

\noindent$\bullet$\textbf{ Limitation of \textit{NeuronZero}.}
 However, we notice that this preliminary attack suffers from the limitation of stealthiness: The dummy neurons can be easily detected due to the anomaly weight distribution. For example, to determine whether the protected models in Fig.\ref{fig:naive_dn} are maliciously injected with dummy neurons, the watermark verifier would first check all the neurons in the suspect model by inspecting the values of incoming and outgoing weights. Once the verification process finds out that the input or output parameters of a certain neuron all equal to value $0$, this neuron is likely to be dummy neurons and is provably has no contribution to the output of the watermarked model. Therefore, the watermark verifier can safely eliminate the detected dummy neurons and the associated weights before the ownership verification.


%% file: tex/attack.tex
\section{Neural Structural Obfuscation}

\subsection{Attack Overview}
In Fig.\ref{fig:atk_pipeline}, our proposed attack consists of three major steps.
1) \textit{Dummy Neuron Generation}: Inspired by the preliminary attack in Section \ref{sec:motivation}, we propose two non-trivial dummy neuron generation primitives, i.e., \textit{NeuronClique} and \textit{NeuronSplit}, to construct dummy neurons associated with non-vanishing weights. This inhibits the direct detection based on the vanishing weights.
2) \textit{Dummy Neuron Injection}: The adversary will generate and inject the dummy neurons from the back to front considering the stealthiness of these injected neurons.
3) \textit{Further Camouflage}: The final step is to further camouflage the dummy neurons in terms of the scale, the location, and the shape of the associated weights via other invariant transforms on DNNs, aiming at transforming the original model into an the obfuscated model which has almost no structural similarity with its original self, while provably preserves the normal model utility. 

Without loss of generality, we present our methodology below with a $H$-layer CNN $f$, where the weights of the $l$-th convolutional layer is denoted as $W^l \in \mathbb R^{N_{l-1} \times N_l \times h \times w}$. Following the notations in Section \ref{sec:motivation}, we denote the input and the output weights of the dummy neuron $m^l_k$ injected in the $l$-th hidden layer as $U^l_{k,in} = \{u^l_{i,k}\}_{i = 1}^{N_{l-1}}$ and $V^l_{k,out} =\{v^{l+1}_{k,j}\}_{j = 1}^{N_{l+1}}$, respectively. We do not consider the bias term because modern DNN models with batch normalization usually have no bias terms \cite{torchvision}. Appendix \ref{sec:app:coverage} provides the technical details on applying our attack framework to more complicated neural architectures (e.g., ResNet \cite{he2016resnet} and Inception \cite{szegedy2016inception}).

\subsection{Dummy Neuron Generation}
\label{sec:dn_generation}
How to construct dummy neurons with non-vanishing input or output weights is challenging. It is mainly because, when the adversary has no knowledge on the input data to the victim model, the contribution of a newly added neuron with non-vanishing weights to the next layer is highly unpredictable, likely to cause a noticeable loss to the original model performance. To eliminate the negative impact of such dummy neurons on the victim model, we alternatively construct groups of \textit{dummy neurons} which work together to preserve the model's prediction behavior. We devise the following structural obfuscation primitives, i.e., \textit{NeuronClique} and \textit{NeuronSplit}. Appendix \ref{sec:app:proof} rigorously proves why these two primitives construct valid groups of dummy neurons.

\begin{figure}[t]
\begin{center}
\includegraphics[width=0.4\textwidth]{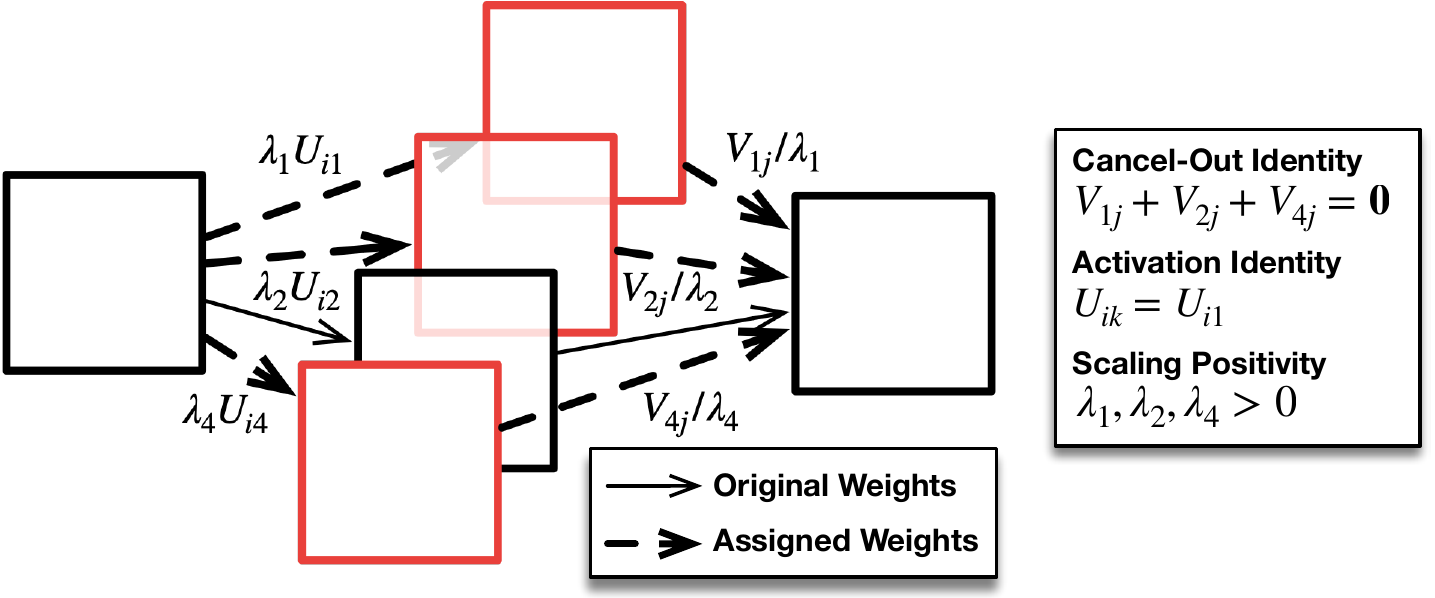}
\caption{A schematic diagram of \textit{NeuronClique} combined with the parameter rescaling invariance.}
\label{fig:neuron_clique}
\end{center}
\vspace{-0.2in}
\end{figure}

\noindent$\bullet$\textbf{ \textit{NeuronClique}.}
Our first proposed primitive, \textit{NeuronClique}, generates a group of dummy neurons assigned with the identical incoming weights and arbitrary outgoing weights which satisfies that they cancel the others' output out. In the following, we elaborate on the case where the adversary attempts to generates $d(\geq 2)$ dummy neurons, i.e., $\{m^l_1,m^l_2,...m^l_d\}$, for the $l$-th layer of the target model $f$ with \textit{NeuronClique}. Formally, the input and output weights of these dummy neurons are designed to satisfy the following conditions:

\begin{equation}
\label{eq:NC_incoming}
U^l_{k,in} = U^l_{1,in} \in \mathbb {R}^{N_{l-1} \times w \times h}, \text{for}\ k = 1,2,...,d,
\end{equation}

\begin{equation}
\label{eq:NC_outgoing}
\sum _{k=1}^d V^l_{k,out} = \textbf{0} \in \mathbb {R}^{N_{l+1} \times w \times h},
\end{equation}
where $U^l_{1,in}$ and $\{V^l_{k,out}\}_{k = 1}^{d-1}$ are randomly sampled from the parameter distribution of the normal neurons to implement by-design stealthiness. For example, in Fig.\ref{fig:neuron_clique}, we generate three dummy neurons by \textit{NeuronClique} and inject them into the $l$-th layer of the prototypical target model. 

\begin{figure}[t]
\begin{center}
\includegraphics[width=0.45\textwidth]{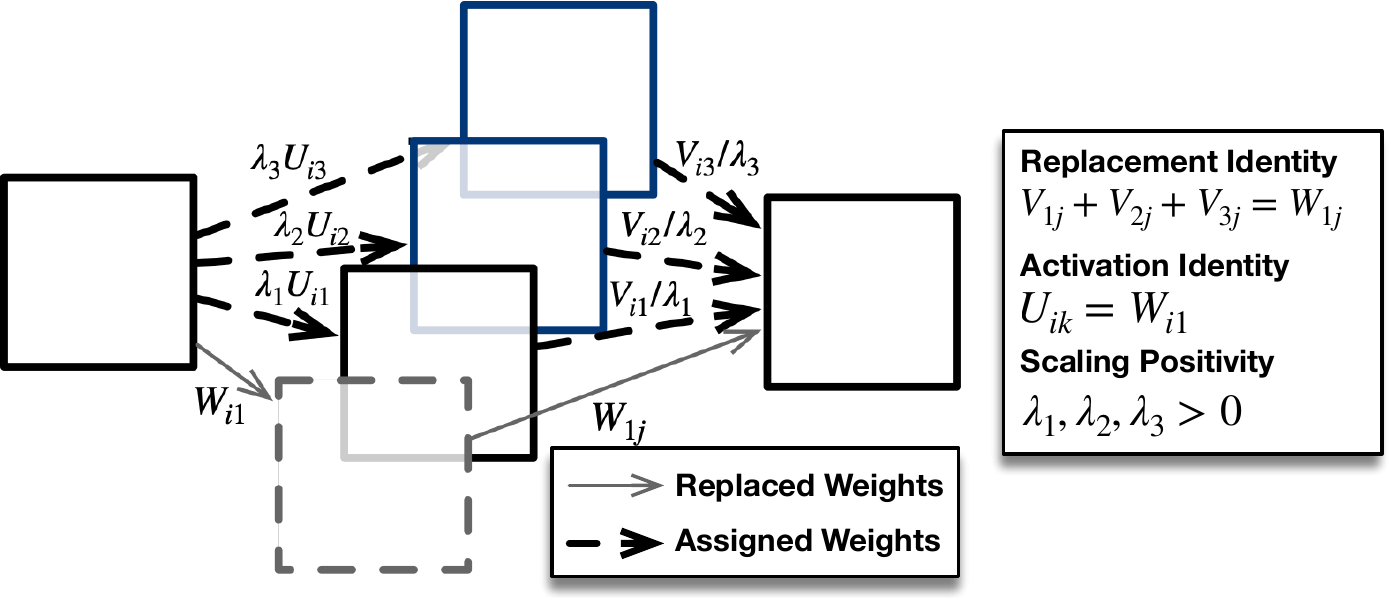}
\caption{A schematic diagram of \textit{NeuronSplit} combined with the parameter rescaling invariance.}
\label{fig:neuron_split}
\end{center}
\vspace{-0.3in}
\end{figure}

\noindent$\bullet$\textbf{ \textit{NeuronSplit}.}
Our second primitive for dummy neuron generation further aims at enhancing the connection between the outputs of the dummy neurons and the original neurons. Specifically, the primitive \textit{NeuronSplit} converts the normal neuron into a number of substitute neurons which work together to preserve functionality of the replaced neuron  for the following layers. Without loss of generality, we split the first neuron in $l$-th layer, i.e., $n^l_1$, into $d+1$ substitute neurons, i.e., $\{m^l_1,m^l_2,m^l_3,...m^l_{d+1}\}$, by which we replace the original neuron with $m^l_1$ and view others as $d$ dummy neurons. Formally, we construct these substitute neurons by setting the input and output weights to satisfy the conditions:
\begin{equation}
\label{eq:Ns_incoming}
U^l_{k,in} = W^l_{1,in} \in \mathbb {R}^{N_{l-1} \times w \times h}, \text{for}\ k = 1,2,...,d+1,
\end{equation}


\begin{equation}
\label{eq:Ns_outgoing}
\sum _{k=1}^{d+1} V^l_{k,out} = W^l_{1,out} \in \mathbb {R}^{N_{l+1} \times w \times h},
\end{equation}
where $W^l_{1,in},W^l_{1,out}$ are the associated incoming weights and outgoing weights of the original neuron $n^l_1$ before the substitution, $\{V^l_{k,out}\}_{k = 1}^{d}$ are randomly sampled from the similar distribution of the normal neurons' weights. For example, as is shown in Fig.\ref{fig:neuron_split}, we can split the first neuron in $l$-th layer into three neurons by \textit{NeuronSplit}. Then we replace the original neuron with one substitute neuron and two dummy neurons into the same layer. 


\subsection{Dummy Neuron Injection}
\label{sec:dn_injection}

Without knowing the specific layers where the watermark is embedded, our attack framework randomly generates and injects groups of dummy neurons via the two primitives (i.e., \textit{NeuronClique} and \textit{NeuronSplit})  \textit{from the last hidden layer to the first layer} of the victim model. Specifically, As is shown in Fig.\ref{fig:atk_pipeline}, we first inject dummy neurons into $l$-th layer and then insert the dummy neurons generated by \textit{NeuronSplit} into $(l-1)$-th layer. It turns out that each dummy neuron in the first group now has different incoming weights with one another, as the these weights of each dummy neuron in $l$-th layer are expanded with the randomly sampled output parameters of the dummy neurons in $(l-1)$-th layer, which does not satisfy Eq.(\ref{eq:NC_incoming}) any longer. This substantially increases the detection budget for a skilled defender in Section \ref{sec:eval:stealthiness_dn}.

As a final remark, generating each dummy neuron group only involves sampling random vectors and several floating-point operations, which incurs almost no additional computational costs than training/finetuning the protected model as in previous attacks. It is worth to note that, the injection process has no change on the dimension of the model's input and output dimension, leaving the prediction API of the victim model in its original form. 



\subsection{Advanced Camouflage Techniques}
\label{sec:dn_camouflage}
To wind up, our proposed attack further camouflages the injected dummy neurons among the original neurons by obfuscating the location, the scale, the shape, and the distribution of the weights associated with the dummy neurons, which are stealthier against possible elimination attempts compared to the preliminary dummy neurons.

\noindent$\bullet$\textbf{ Exploiting Shuffling and Scaling Invariance.}
We offensively exploit the usage of the shuffling and scaling invariance in DNN \cite{neyshabur2015pathsgd, ganju2018property}. With the scaling invariance, the adversary can obfuscate the weight scale by scaling up the incoming weights of every injected dummy neuron with a positive value $\lambda$ and scaling down its outgoing weights at the same ratio, which can prevent the weight distribution of the dummy neurons from being abnormal as these weights in scaling equivalence do not satisfy the conditions of \textit{NeuronClique} or \textit{NeuronSplit} any longer. 
To randomize the fake weight's location, our attack framework leverages the permutation invariance of neural networks to disperse the dummy neurons among the original ones. We use random permutation on the neurons of every expanded layer to randomize the location of the injected dummy neurons, rather than injecting the dummy neurons as an independent module, or otherwise the location information would be exploited for neuron inspection.

\noindent$\bullet$\textbf{ Kernel Expansion.}
Furthermore, the adversary can further modify the architecture of the protected model by expanding the kernel size of every convolutional layers, which obfuscates the weight shape of the dummy neurons. For intuition, we can obtain an equivalent model by padding the kernel matrix with all $0$, while increasing the amount of the implicit padding of the input activation maps. Combined with the injected dummy neurons, our proposed attack can pad the kernels within the same layer with non-vanishing values, which not only improves the stealthiness of injected neurons but also introduces more perturbation to the verification process. More technical details on kernel expansion are in Appendix \ref{sec:app:kernel_expansion}. 


%% file: tex/evaluation.tex
\section{Evaluation and Results}
\label{sec:eval}
 \noindent\textbf{Overview of Evaluation.} To validate the effectiveness of our attack, we provide a systematic study on the vulnerability of nine state-of-the-art white-box watermark schemes published at top-tier venues and span the different stages of DNN watermark development.
 Before the detailed evaluation results, we concisely introduce the evaluation setups.

\input{tex/tables/new_scenarios.tex}


\noindent$\bullet$\textbf{ Target Watermark Schemes.} As Table \ref{tab:new_scenario} shows, our evaluation covers most of the existing white-box DNN watermarks spanning the different stages of DNN watermark development, which faithfully reflected the three representative branches, i.e., \textit{weight-based, activation-based and passport-based} schemes.
\begin{itemize}
    \item \textit{Weight-based Watermarks \cite{uchida2017embedding,wang2021riga,liu2021greedyresiduals,ong2021iprgan}}: In weight-based watermarking schemes, the legitimate owner would secretly select one/more neural layers from the target model, and embed the identity message into a preprocessed version of the layer parameters during the training process.    
    \item \textit{Activation-based Watermarks \cite{darvish2019deepsigns,lim2022ipcaption}}: In contrast to the weight-based ones, activation-based watermarking schemes embed the identity message into the activation maps of a special set of data inputs at the target layers. Some works argue activation-based watermarks are more dynamic and data-dependent, which makes them more robust against obfuscation on model parameters\cite{darvish2019deepsigns}.  
    \item \textit{Passport-based Watermarks \cite{zhang2020passportaware,fan2021deepip}}: Passport-based watermarking schemes can be viewed as a hybrid of the weight-based and the activation-based schemes. On the one hand, the identity information is typically embedded to the learnable parameters of normalization layers in the target model. On the other hand, they reassert the validity of the ownership by inserting a special passport layer into the suspect model to check whether the DNN model inference performance is unyielding \cite{fan2021deepip}.
\end{itemize}

\noindent$\bullet$\textbf{ Choices of Error-Handling Mechanisms.} We find most of the tested watermark schemes in Table \ref{tab:new_scenario} (except Greedy Residuals \cite{liu2021greedyresiduals}) are not executable when the parameters or the activation maps from the target layers are incompatible with the shape constraint of the predefined watermark extraction algorithm $\mathcal{E}$. To evaluate watermark existence, we therefore propose to implement an error-handling mechanism, which restores watermark-related parameters/activation maps to a valid size. Specifically, we get inspirations from \cite{liu2021greedyresiduals} to greedily remove the neurons with the less absolute mean value of the incoming and outgoing weights for each layer before the model watermark extraction. We call this error-handling strategy as \textit{Max-First}. In Section \ref{sec:eval:stealthiness_dn}, we further investigate more adaptive defenses on dummy neurons.


\noindent$\bullet$\textbf{ Implementation of Watermark Schemes.} For each watermarking scheme, we strictly follow the same experiment settings in the official implementations to reproduce a watermarked model for fair evaluation. This includes but not limited to the model architecture, dataset and watermark-related hyper-parameters, on which they claim the robustness to existing removal attacks. Also, we employ the same signature $s=$``\textit{this is my signature}'' in these known watermark schemes. These methods protect the IP of diverse models, including ResNet, Inception for image classification, DCGAN for image generation and LSTM for image captioning task, which hence support the broad applicability of our proposed attack. 

\noindent$\bullet$\textbf{ Evaluation Metrics.} For attack effectiveness, we use \textit{Bit Error Rate} (BER), i.e.,  the proportion of modified bits in the extracted watermark compared to the pre-defined signature,  to measure how much the watermark is tampered by our removal attack. To compare the robustness of different white-box watermark schemes we follow \cite{sokwatermark} to determine the decision threshold and then re-scale the BER results. Table \ref{tab:new_scenario} summarizes the decision threshold for the mainstream white-box watermark schemes. After modeling the decision threshold for each watermark, we also define a linear scaling function $S(x, \theta)$ similar to \cite{sokwatermark} as follows: $S(x ; \theta)=\min \left(1, \frac{\theta^{\prime}}{\theta} x\right)$,
which relates the BER from different white-box watermark methods with each other. As a result, the watermark is removed if the rescaled BER is higher than $50\%$. Otherwise, the watermark is retained. For utility loss, we report the performance of the watermarked model before/after our removal attack, i.e., FID \cite{heusel2017FID} for image generation and BLEU-1 \cite{Papineni2002BleuAM} for image captioning task and classification accuracy for other tasks \cite{ong2021iprgan,lim2022ipcaption}.


\begin{figure*}[h]
\begin{center}
\includegraphics[width=1\textwidth]{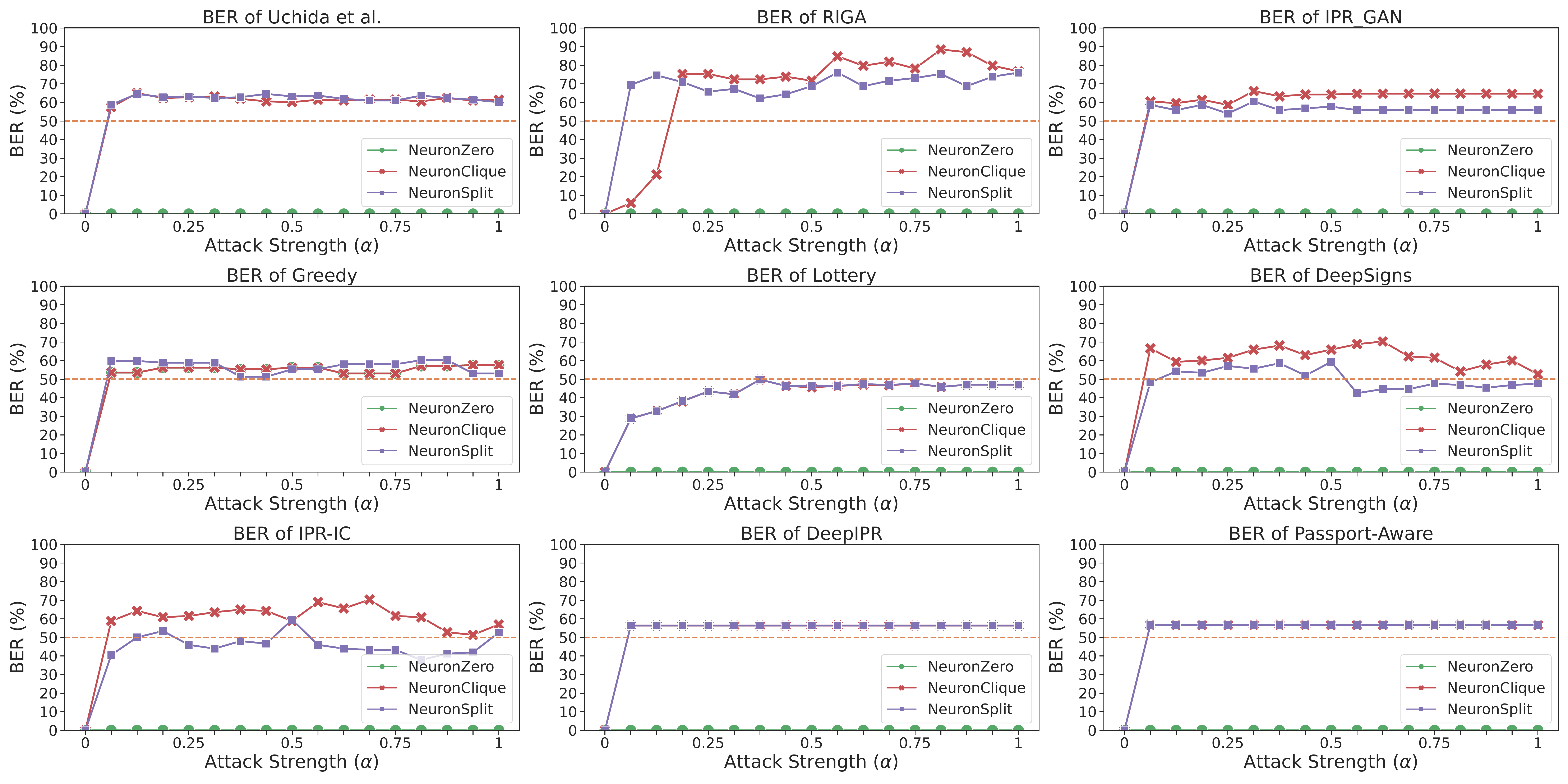}
\vspace{-0.3in}
\caption{The re-scaled BER of the watermarked models after inserting a certain proportion of neurons by our attacks. The dashed horizontal lines report the BER of the irrelevant mode.}
\label{fig:scaled_ber}
\end{center}
\vspace{-0.3in}
\end{figure*}


\noindent$\bullet$\textbf{ Summary of Results.} 
 Fig.\ref{fig:scaled_ber} summarize the effective of our proposed attacks, where the $x$-axis reflects the \textit{attack strength} defined as the ratio of added and modified dummy neurons in the target neural network.  We make the observation that most of the resulting scaled BER exceeds the removal threshold 50\%.  This validates that, due to the heavy assumption of existing schemes on the integrity of the neural architecture, most of them lose the claimed robustness to some or all the previously known attacks when evaluated under our proposed attack. 
 To apply error-handling mechanisms such as Max-First cannot restore the embedded watermark from structural obfuscated model by NeuronClique and NeuronSplit. In most cases, the BER is increased over 50\% once we add less than 5\% dummy neurons, indicating the innate vulnerability of these white-box watermarking schemes.  In the following, we provide a case-by-case analysis on the vulnerability of each scheme and how our attack cracks them by neural structure obfuscation with dummy neurons. 

 

\input{tex/cases.tex}
\input{tex/defense.tex}

%% file: tex/tables/new_scenarios.tex

\begin{table}[t]
  \centering
  \caption{
  Summary of nine mainstream white-box model watermarks embedded in industry-level DNNs in our experiments. Results in the \textbf{Utility} column report the performance of the protected model before/after our attack.}
  \scalebox{0.65}{
    \begin{tabular}{llllc}
    \toprule
 {\textbf{Category}}  & {\textbf{Scheme}}  & {\textbf{Model}}  & {\textbf{Utility}} & \textbf{Threshold} \\
     \midrule
    \multirow{5}{*}{Weight} & Uchida et al. \cite{uchida2017embedding} & WRN  & 91.55\%/91.55\% & 43.86\% \\
      & RIGA \cite{wang2021riga} & Inceptionv3  & 95.90\%/95.90\% & 42.74\%  \\
      & IPR-GAN \cite{ong2021iprgan} & DCGAN & 54.33/54.33 & 41.96\%  \\
      & Greedy \cite{liu2021greedyresiduals} & ResNet18  & 55.05\%/55.05\%  & $43.77\%$ \\
      & Lottery  \cite{chen2021lottery} & ResNet18 & 66.40\%/66.40\% & 34.50\% \\
    \midrule
    \multirow{2}{*}{Activation} & DeepSigns \cite{darvish2019deepsigns} & WRN  & 89.94\%/89.94\% & 42.68\%  \\
       & IPR-IC \cite{lim2022ipcaption} & ResNet50+LSTM & 72.06/72.06 & 46.22\% \\
    \midrule
    \multirow{2}{*}{Passport} & DeepIPR \cite{fan2021deepip} & ResNet18 & 67.94\%/67.94\% & 46.26\% \\
         & Passport-Aware \cite{zhang2020passportaware} & ResNet18 & 74.78\%/74.78\% & 45.79\% \\
    \bottomrule
    \end{tabular}}%
  \label{tab:new_scenario}%
  \vspace{-0.1in}
\end{table}%

%% file: tex/cases.tex
\subsection{Attacking Weight-based Watermarks}
\label{sec:eval_weight}
\noindent\textbf{(1) Uchida et al.}
Uchida et al.\cite{uchida2017embedding} introduce one of the earliest white-box schemes which embed the model watermark into the convolutional weights of the target model. 
To extract the model watermark, the scheme first averages the convolutional weights $w \in \mathbb{R}^{N_{l-1}\times N_l \times w \times h}$ of the watermark-related layer through first dimension and flattens the weight to $\hat w \in \mathbb{R}^{(N_l \cdot w \cdot h)}$. Then, a binary string $s'$ is obtained from $\hat w$ through a pre-defined linear matrix $A$ and a threshold function $T_h$ at 0, i.e.,
$
s' = T_h(X \cdot \hat w),
$
which is matched with the owner-specific message $s$ in terms of BER for validation.

\noindent$\bullet$\textbf{ Discussion.}
Although this approach is previously shown to be vulnerable to overwriting, known attacks however require the specific prior knowledge of the watermark algorithm as well as a domain dataset, both of which are usually impractical. Our attack reveals the insecurity of this algorithm by directly modifying the structure of the target model while leaving the utility intact. Specifically, the dimension of $\hat w$ is increased once the adversary injects dummy neurons into watermark-related layers. As a result, during the extraction procedure of $s_i'$ from the victim model, the second dimension of pre-defined linear transformation matrix $X$ is unmatched to the dimension of expanded $\hat w$ any longer.


\noindent$\bullet$\textbf{ Evaluation Results.}
We run the code of \cite{code-uchida} they publicly released to reproduce a watermarked model of Wide Residual Network (WRN) trained on CIFAR10 dataset. We perform our removal attack to inject dummy neurons generated via different methods into each layer, which presents the same original utility with classification accuracy of $91.55\%$ while the verification procedure is inhibited if with no error handling mechanism. As Fig.\ref{fig:scaled_ber} shows that applying Max-First cannot restore the embedded watermark from structural obfuscated model by \textit{NeuronClique} and \textit{NeuronSplit}, as the BER is increased over $50\%$ once we add less than $5\%$ dummy neurons, indicating the innate vulnerability of this scheme.

\noindent\textbf{(2) RIGA.} Wang et al. \cite{wang2021riga} replace the linear transformation matrix in Uchida et al. with a learnable fully-connected neural network (FCN) to boost the encoding capacity of watermarking messages. That is, the intuition behind this watermark scheme is closely identical to Uchida et al. We present the full details in Appendix \ref{sec:app:eval} and report the results in Fig.\ref{fig:scaled_ber}.
\noindent\textbf{(3) IPR-GAN.}
To extend the watermarking primitive to generative adversarial networks (GANs) \cite{goodfellow2014gan}, Ong et al. present the first model watermark framework which first invokes black-box verification to collect some evidence from the suspect model via remote queries, and then utilizes the white-box verification for further extracting the watermark from the specific weights of suspicious model through the law enforcement. Different from \cite{uchida2017embedding}, Ong et al. embed the identification information into the scale parameters $\gamma$ of the normalization layers, rather than the convolutional weights. Correspondingly, the transformation function used in watermark verification stage consists of only a hard threshold function $T_h$, which actually extracts the signs of $\gamma$ in selected normalization layers as a binary string, i.e., 
$s' = T_h(\gamma).$

\noindent$\bullet$\textbf{ Discussion.}
We focus on the white-box part of the watermark method. Previous works have shown that the scale parameters $\gamma$ of normalization layers are more stable than the convolution weights against existing removal attacks and ambiguity attacks, as small perturbation to these watermark-related parameters would cause significant drops to the original model utility. However, the number of scale weights in the watermark-related layer can be increased once we inject a group of dummy neurons. As a result, the length of binary string $s'$ extracted by the transformation function $T_h$ in this watermarking scheme is incompatible with the length of the target watermark any longer.

\noindent$\bullet$\textbf{ Evaluation Results.}
We follow their evaluation setups to watermark DCGAN trained on the CUB200 dataset, which achieves $54.33$ in terms of FID and has $0\%$ BER \cite{code-iprgan}.
As the watermark verification procedure is blocked with our proposed removal attacks, we leverage the error-handling methods on the scale weights to measure the BER of the extracted signature, which only has at most $55.86\%$ matched bits to the pre-defined binary signature while the FID of the generator is perfectly preserved as $54.33$, as  Fig.\ref{fig:scaled_ber} shows.

\noindent\textbf{(4) Greedy.}
Liu et al. \cite{liu2021greedyresiduals} propose to greedily select fewer yet more important model weights called the \textit{greedy residuals}, which is more important to the normal model utility and hence improves the watermark robustness against previous attacks. Specifically, the method extracts the identity information by first applying the transformation function $A$ on the one-dimensional average pooling over the flattened parameters $\hat w$ in the chosen convolutional layers, and then greedily takes the largest absolute values in each row by a ratio of $\eta$ to build the residuals. Finally, the secret binary string can be extracted from the signs of residuals by hard threshold function $T_h$ after being averaged to a real-valued vector.
Formally, the extraction procedure can be written as
$
s' = T_h(Avg(Greedy(Avg\_pool\_1D(\hat w)))).
$

\noindent$\bullet$\textbf{ Discussion.}
Greedy Residuals utilize some important parameters for embedding, which are more stable than choosing all the convolution weights in the specific layer proved in their ablation evaluations. Moreover, this watermark scheme greedily select the larger absolute values in each row from the intermediate feature matrix to build the residuals with fixed number of values, the verification procedure is not inhibited with the injection of dummy neurons. However, as the average pooled feature matrix before residual construction is impacted by some arbitrary values introduced by the injected dummy neurons, the extracted watermark after our attack would be perturbed unexpectedly. 

\noindent$\bullet$\textbf{ Evaluation Results.}
We run the source codes of Greedy Residuals publicly released by the authors \cite{code-greedy} to reproduce a watermarked ResNet18 training on Caltech256 dataset with $55.05\%$ accuracy and $0\%$ BER. We embed the secret watermark on the parameters of the first convolution layer with $\eta = 0.5$. We prove that our removal attack can utterly destroy the model watermark embedded into the residual of fewer parameters, for example, leading to an increase in the BER to $57.57\%$ after injecting dummy neurons from \textit{NeuronClique} whereas the model utility remains unchanged.

\noindent\textbf{(5) Lottery.}
The Lottery Ticket Hypothesis (LTH) explores a new scheme for compressing the full model to reduce the training and inference costs. As the topological information of a found sparse sub-network (i.e., the winning ticket) is a valuable asset to the owners, Chen et al. propose a watermark framework to protect the IP of these sub-networks \cite{chen2021lottery}. Specifically, they take the structural property of the original model into account for ownership verification via embedding the watermark into the weight mask in several layers with highest similarity to enforce the sparsity masks to follow certain 0-1 pattern. The proposed lottery verification uses the QR code to increase the capacity of the watermark method. For watermark verification, this algorithm first selects a set of watermark-related weight masks $m$ and then averages the chosen masks to a 2-dimensional matrix, which is further transformed to a QR code via $Sign$ function, i.e., $s' = Sign(Avg(m))$ and can be validated via a QR scanner. 

\noindent$\bullet$\textbf{ Discussion.}
While most existing watermark techniques explore the specific model weights or prediction to embed the secret watermark, the lottery verification leverages the sparse topological information (i.e., the weight masks) to protect the winning ticket by embedding a QR code which can be further decoded into the secret watermark. However, our attack directly obfuscates the topological information of the target model by injecting groups of dummy neurons with the corresponding weight masks, which enlarges the shape of extracted QR code from the weight mask of trained winning ticket unexpectedly. As this QR code without valid shape is not decodable, we leverage the error-handling mechanism to extract the embedded watermark for ownership verification, where remains a large distortion.

\begin{figure}[t]
\begin{center}
\includegraphics[width=0.4\textwidth]{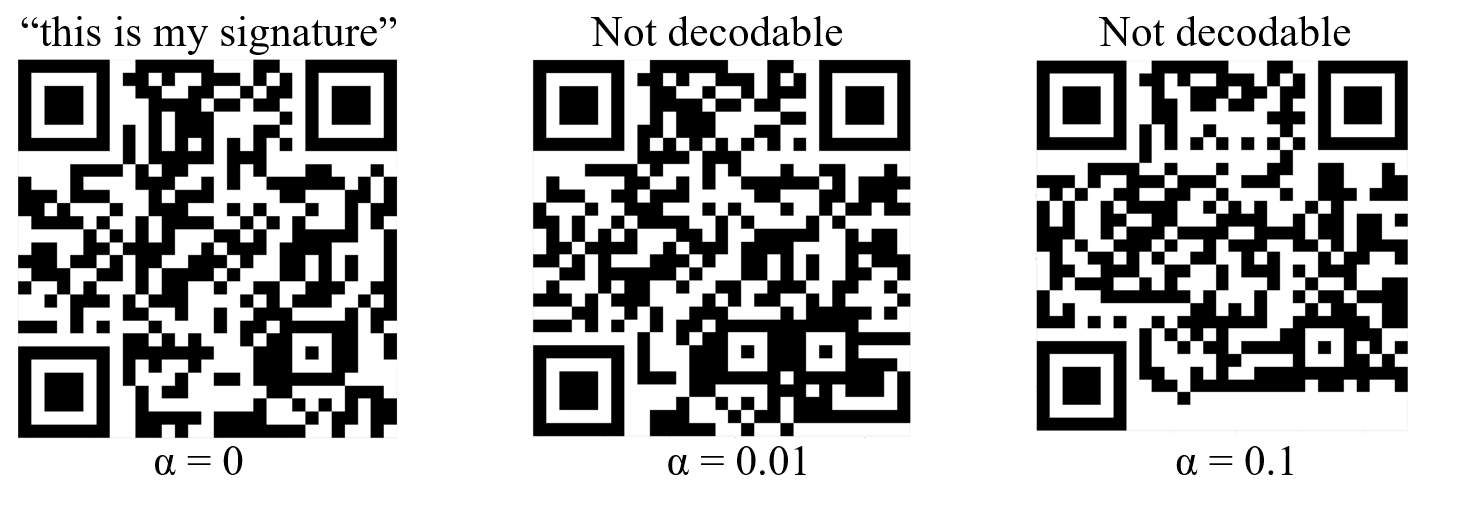}
\vspace{-0.2in}
\caption{The QR codes extracted from ResNet-20 watermarked by\cite{chen2021lottery} after an $\alpha$ ratio of dummy neurons are added.}
\label{fig:qrcode}
\end{center}
\vspace{-0.3in}
\end{figure}

\noindent$\bullet$\textbf{ Evaluation Results.}
We follow the evaluation settings in the original paper to watermark a ResNet20 training on CIFAR-100 dataset, which achieves $66.41\%$ accuracy and $0\%$ BER \cite{code-lottery}. Although the QR code has the ability to correct some errors which improves the robustness against existing attacks, the identity information decoding procedure from the extracted QR code is invalidated by adding only a few (e.g., $1\%$)  dummy neurons in the victim model as Fig.\ref{fig:qrcode} shows. Moreover, the original design of Lottery Ticket is inhibited (due to the unmatched size) when attackers insert the dummy neurons into the protected model, while it retains robust against structure obfuscation with our proposed error-handling mechanisms. In other words, Lottery Ticket would have better robustness against neural structural obfuscation than other schemes if the unmatched size of neural layers are properly addressed in its design. 

\subsection{Attacking Activation-based Watermarks}
\noindent\textbf{(1) DeepSigns.}
As a representative of activation-based white-box watermarking schemes, DeepSigns proposes to embed the model watermark into the probability density function (PDF) of the intermediate activation maps obtained in different layers on the white-box scenario. Specifically, DeepSigns adopts a Gaussian Mixture Model (GMM) as the prior probability to characterize the hidden representations, and considers the mean values of the PDF at specific layers to share the same role as the watermark-related weights in Uchida et al. \cite{uchida2017embedding}. Similar to the verification procedure of \cite{uchida2017embedding}, a transformation matrix $A$, randomly sampled in embedding procedure, projects the mean values of chosen intermediate features to a real-valued vector. With the final hard threshold function, the resulted binary string $s'$ is matched to the owner-specific watermark for claiming the model ownership. 


\noindent$\bullet$\textbf{ Discussion.}
The notable difference between DeepSigns and \cite{uchida2017embedding} is where to embed the model watermark. 
However, as the hidden features utilized by DeepSigns are generated by the weights in the preceding layer, e.g., $a_i = W_i\cdot x+b_i$, the structural information of target model is closely related to the shape of output feature maps. 
For example, the shape of the watermark-related layer's output is expanded after injecting dummy neurons, which inhibits the ownership verification due to the incompatible dimension of the output activation maps and the pre-defined transform matrix.
As a result, DeepSigns is almost as vulnerable as \cite{uchida2017embedding} under our attack.

\noindent$\bullet$\textbf{ Evaluation Results.}
We run the source code of DeepSigns from \cite{code-deepsign} to watermark a wide residual network trained on CIFAR10. This watermarked WRN achieves $89.94\%$ accuracy and $0\%$ BER. With the error-handling method, it is shown that the ownership verification of the target model is completely confused by our removal attacks. For example, the BER is increased to $47.59\%$ with dummy neurons from \textit{NeuronSplit} while the original model functionality is intact.

\noindent\textbf{(2) IPR-IC.}
As previous watermarking schemes are mainly designed for image classification models, they are insufficient to IP protection for image captioning models and cause inevitable degradation to the image captioning performance. Therefore, Lim et al. \cite{lim2022ipcaption} embed a unique signature into Recurrent Neural Network (RNN) through hidden features. At the ownership verification stage, the mask matrix $M$ first selects the hidden memory state $h$ of given watermarking image in protected RNN model. Then, the hard threshold function transforms the chosen $h$ to a binary string $s'$, which can be formally written as
$
s' = T_h(h).$

\noindent$\bullet$\textbf{ Discussion.}
Similar to DeepSigns \cite{darvish2019deepsigns} and IPR protection on GANs \cite{ong2021iprgan}, the topological information is closely related to the shape of hidden memory state. Although the protected image captioning model contains an RNN architecture, we can adopt our structural obfuscation method to expand the size of hidden state, e.g., with vanishing weights, to produce an equivalence of the original model with the same output.

\noindent$\bullet$\textbf{ Evaluation Results.}
We run the official implementation \cite{code-captioning} to reproduce a watermarked Resnet50-LSTM trained on COCO, which achieves $72.06$ BLEU-1 and has $0\%$ BER. With our proposed removal attacks, the signature extracted from the hidden memory state $h$ is not compatible to the size of the owner-specific binary message. We leverage error-handling mechanisms, e.g., Max-First to extract the identity message, which is perturbed with $56.95\%$ and $52.60\%$ BER for \textit{NeuronClique} and \textit{NeuronSplit}, while no loss is brought to the image captioning performance.

\subsection{Attacking Passport-based Watermarks} 
\noindent\textbf{(1) DeepIPR.} 
DeepIPR is one of the earliest passport-based DNN ownership verification schemes \cite{fan2021deepip}. By inserting owner-specific passport layers during the watermark embedding procedure, DeepIPR is designed to claim the ownership not only based on the extracted signature from the specific model parameters but also on the model performance with the private passport layer. Consequently, this scheme shows high robustness to previous removal attacks and especially to the ambiguity attacks, which mainly forge counterfeit watermarks to cast doubts on the ownership verification.

In our evaluation, we focus on the following passport verification scheme in \cite{fan2021deepip}. This scheme generates two types of passport layers simultaneously by performing a multi-task learning, i.e., public passports for distribution and private passports for verification, both of which are actually based on normalization layers. Generally, DeepIPR leverages pre-defined digital passports $P=\{P_{\gamma}, P_{\beta}\}$ to obtain the scale and the bias parameters of the private passport, which are written as:
$
\gamma = Avg(W_c \odot P_{\gamma}), \beta =  Avg(W_c \odot P_{\beta}),
$
where $W_c$ is the filters of the precedent convolution layer, and $\odot$ denotes the convolution operation. DeepIPR adopts a similar watermark extraction process as \cite{uchida2017embedding}, where the transformation function $A$ converts the signs of the private $\gamma$ into a binary string to match the target signature.

\noindent$\bullet$\textbf{ Discussion.}
As the private $\gamma$ and $\beta$ are obtained from the preceding convolution weights, DeepIPR actually embeds the secret signature in the hidden output of the convolutional layer with the weights $W_c$ given the input $P_{\gamma}$ or $P_{\beta}$. Similar to the activation-based watermarking scheme, the unmatched extracted watermark can not be used to ownership verification due to the expanded shape of $W_c$.


\noindent$\bullet$\textbf{ Evaluation Results.}
We evaluate our attack on the watermarked ResNet18 trained on the CIFAR-100 dataset with DeepIPR \cite{code-deepipr} which achieves $67.94\%$ accuracy. When we inject an $\alpha$ proportion of neurons with our attack, the signature extracted from the victim model becomes totally unreadable, from \textit{``this is my signature''} ($\alpha =0$, BER$=0\%$) to \textit{``ÎÍ¿±C¾Ýzü½¤L°!²/Ã9Ûå''} ($\alpha=0.5$, BER$=51.25\%$). By injecting $50\%$ of dummy neurons, our attack
successfully increases the BER to almost random, while causing no change in the accuracy of the model with the public passports. 

\noindent\textbf{(2) Passport-aware Normalization.} Zhang et al. \cite{zhang2020passportaware} propose another passport-based watermark method without modifying the target network structure by maintaining the statistic values independently for passport layer. As the watermark embedding and extracting procedures are nearly identical to DeepIPR, we report the results in  Fig.\ref{fig:scaled_ber} and provide the details of Passport-aware Normalization in Appendix \ref{sec:app:eval}.

%% file: tex/defense.tex
\subsection{Stealthiness of Injected Dummy Neurons}
\label{sec:eval:stealthiness_dn}
Finally, we provide a preliminary study on potential adaptive approaches to detect and eliminate dummy neurons. Specifically, we consider two types of defenders, (a) a \textit{partially knowledgeable defender}, i.e., who knows the existence of dummy neurons but has rare knowledge about the detailed algorithm for generating the dummy neurons, (b) a \textit{skilled defender}, who has a perfect knowledge about our attack framework but does not have access to the original model (a common setting in existing watermarking protocols), and (c) a \textit{fully knowledgeable defender}, who also has the original model for reference. 

\noindent\textbf{(a) A Partially Knowledgeable Defender.} If knowing the existence of dummy neurons in the suspect model, the defender is likely to apply anomaly detection algorithms to detect and eliminate the suspicious neurons from the target layer. Specifically, by considering the incoming and outgoing weights of each neuron as the feature vector, we implement two representative anomaly detection algorithms, i.e., \textit{cluster-based} \cite{chen2018activation_clustering} and \textit{SVD-based} \cite{tran2018spectral}, to evaluate the stealthiness. 

\noindent$\bullet$\textbf{ Experimental Settings.}
We first inject the dummy neurons generated by NeuronZero, NeuronClique, and NeuronSplit into the watermarked models, respectively. Then, we concatenate the flattened incoming and outgoing weights of each neuron as its feature vector. The cluster-based detection leverages K-Means to separate the neurons from the same layer and assigns the abnormal cluster as dummy neurons \cite{chen2018activation_clustering}, while the SVD-based detection utilizes the covariance matrix of the neurons' feature representation to filter outliers \cite{tran2018spectral}.

\begin{figure}[t]
\begin{center}
\includegraphics[width=0.5\textwidth]{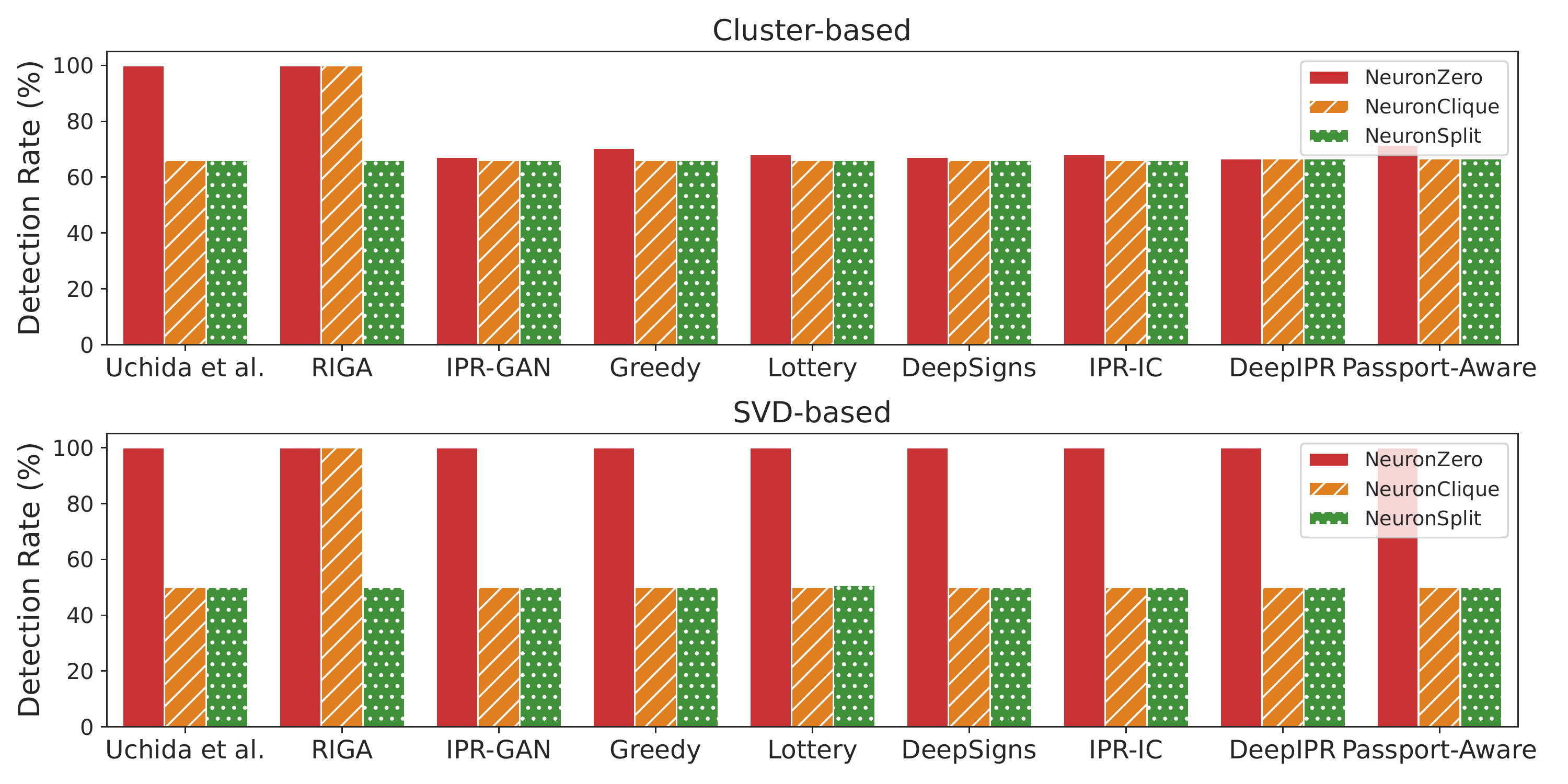}
\caption{Detection rate of anomaly detection algorithms on different types of dummy neurons.}
\label{fig:detection}
\end{center}
\end{figure}

\noindent$\bullet$\textbf{ Results \& Analysis.} 
As is shown in Fig.\ref{fig:detection}, the dummy neurons with vanishing values generated by NeuronZero are more likely to be recognized as abnormal neurons under both detection approaches, while the dummy neurons produced by NeuronSplit from the original neurons show stronger stealthiness compared to both NeuronZero and NeuronClique, as their weights have the same distribution to the normal ones.

\input{tex/algorithm/dn_elimination.tex}

\noindent\textbf{(b) A Skilled Defender.} Besides the above defense, we consider a more adaptive defender who has perfect knowledge about our proposed attack. From our construction in Section \ref{sec:dn_generation} (combined with the defense results above), the only exploitable information for dummy neuron elimination is in the first layer where the dummy neurons are injected. According to Eq.\ref{eq:NC_incoming}\&\ref{eq:Ns_incoming}, if there are no dummy neurons in the previous layer, then the dummy neurons belonging to the same group generated by \textit{NeuronClique} or \textit{NeuronSplit} would have their incoming weights, if viewed as vectors, proportional to each other. Based on this characteristic, the defender may implement the following procedures to detect and eliminate dummy neurons from each layer:
\begin{itemize}
    \item \textit{Step 1.} Normalize the incoming weights of each neuron in the current layer and move the neurons with the same normalized weight into the same hash bucket.
    \item \textit{Step 2.} Merge the neurons of the same hash bucket into one neuron: Its incoming weights take the normalized weights of either one of the neurons and its outgoing weights take the sum of these neurons.  
    \item \textit{Step 3.} After the merging, check the flattened outgoing weights of each neuron: If the weights are a zero vector, then remove the neuron and its associated weights. 
\end{itemize}


  By iterating the above procedure from the first hidden layer to the last one, the algorithm is expected to detect the dummy neurons and restore the original neural architecture from the obfuscated model. The detailed algorithm is shown in Algorithm \ref{alg:dn_elim}. Our experiments find, even though there is no dummy neurons after the elimination, the BER of the recognized watermark in the restored model remain over $50\%$ in Table \ref{tab:elim_table}, yielding no evidence for the claimed ownership. This is because, after the merging, the original scale of the parameter could still not be recovered, because the defender does not know how the attacker has rescaled the parameters of the dummy neurons during NeuronClique and NeuronSplit. Therefore, when the defender has no access to the original model, they could not adjust the parameter scale to cancel out the obfuscation effect. Hence, the original watermark is not recovered.

\input{tex/tables/elim_table}
\noindent\textbf{(c) A Fully Knowledgeable Defender.} Finally, we discuss the case when the defender also refers to the original watermarked model for watermark recovery. Then they can further compare the parameters of the original model with the obfuscated model in order to recover the order and the scale of the parameters. For the dummy neurons constructed in Section \ref{sec:dn_generation}, they can eliminate the dummy neurons and recover the watermark accuracy at the cost of some additional computing power. As the security research on white-box model watermark is an evolving game between the attacks and the defenses, we leave the study on more effective de-obfuscation approaches to future work.

%% file: tex/algorithm/dn_elimination.tex
 \begin{algorithm}[h]
 \caption{A possible dummy neuron elimination algorithm.}
 \label{alg:dn_elim}
  {\fontsize{10}{10}\selectfont
 \begin{algorithmic}[1]
 \renewcommand{\algorithmicrequire}{\textbf{Input:}}
 \REQUIRE $W$ (the parameters of the suspect model), $H$ (the number of layers in the suspect model).
 \renewcommand{\algorithmicensure}{\textbf{Output:}}
 \ENSURE $W$, the parameters of the suspect model after dummy neuron elimination.
 \STATE {$T_{hash} \gets \{\} $}\\
 {\small{/* Find the neurons with proportional incoming weights.*/}}
 \FOR {$l = 1,2,...,H-1$}
  \STATE {$W^{(l)}, W^{(l+1)} \gets W[l],W[l+1]$} {\small{\hskip3em // Obtain the incoming and outgoing weights of the neurons in the $l^\text{th}$ layer}} 
  \STATE {$Ind, \tilde{W}^{(l)}, \tilde{W}^{(l+1)} \gets 0, zeros\_{like}(W^{(l)}), zeros\_{like}(W^{(l+1)})$}

  \FOR {each input weight $W^{(l)}_{\cdot i}$ of the $i^{th}$ neuron in the $l^{th}$ layer}
   \STATE{$w \gets \text{Flatten}(W^{(l)}_{\cdot i}) $}
   \STATE {$w_{norm} \gets \text{Normalize}(W^{(l)}_{\cdot i})$}
   \STATE {$w \gets \frac{W^{(l)}_{\cdot i}}{w_{norm}}$}
   \IF {$w$ not in $T_{hash}$.keys()}
    \STATE {$T_{hash}[w] \gets Ind$}
    \STATE {$Ind \gets Ind + 1$}
    \ELSE
     \STATE{\small{/* Merge the neurons in the same hash bucket.*/}} 
     \STATE {$i' \gets T_{hash}[w]$}
     \STATE {$\tilde{W}^{(l)}_{\cdot i'} \gets w$}
     \STATE {$\tilde{W}^{(l+1)}_{i' \cdot } \gets \tilde{W}^{(l+1)}_{i' \cdot } + w_{norm} \cdot W^{(l+1)}_{i \cdot}$}
    \ENDIF
   \ENDFOR
  \STATE {$W^{(l)}, W^{(l+1)} \gets \tilde{W}^{(l)}, \tilde{W}^{(l+1)}$}
\STATE {Remove the neurons with zero incoming or outgoing weights in $W^{(l)}$, $W^{(l+1)}$.}
  \ENDFOR
 \RETURN  $W$
 \end{algorithmic}}
 \end{algorithm}

%% file: tex/tables/elim_table.tex
\begin{table}[t]
  \centering
  \caption{The scaled BER for each white-box watermarking scheme under dummy neuron elimination.}
  \scalebox{0.85}{
    \begin{tabular}{ccccc}
    \toprule
         \multicolumn{1}{c}{\textbf{Schemes}} & \multicolumn{1}{c}{Uchida et al.} & \multicolumn{1}{c}{RIGA} & \multicolumn{1}{c}{IPR-GAN} & \multicolumn{1}{c}{Greedy} \\
    \midrule
    \multicolumn{1}{l}{\textbf{BER}} & \textbf{52.99\%} & \textbf{54.83\%} & \textbf{62.37\%} & \textbf{51.79\%} \\
    \midrule
    \multicolumn{1}{c}{Lottery} & \multicolumn{1}{c}{DeepSigns} & \multicolumn{1}{c}{IPR-IC} & \multicolumn{1}{c}{DeepIPR} & \multicolumn{1}{c}{Passport-Aware} \\
    \midrule
    \textbf{54.45\%} & \textbf{52.74\%} & \textbf{53.76\%} & \textbf{57.42\%} & \textbf{54.59\%} \\
    \bottomrule
    \end{tabular}}%
  \label{tab:elim_table}%
\end{table}%

%% file: tex/discussion.tex
\section{Discussions}\label{sec:discussion}


\noindent$\bullet$\textbf{ Applicability of Our Attack.}
For simplicity, our methodology part uses the convolutional and fully-connected layers as the motivating example. Nevertheless, First, our proposed neural structure obfuscation with dummy neurons is applicable to many other neural network components, most of which are already involved in the covered victim models we evaluate in Section \ref{sec:eval}. We leave the technical details in Appendix \ref{sec:app:coverage}.

\noindent$\bullet$\textbf{ Coverage of Attack Targets.} Recently, an increasing number of white-box model watermarking schemes have been published at top-tier venues, some of which are authored by industrial research institutes (e.g., from Microsoft \cite{darvish2019deepsigns} and WeBank \cite{fan2021deepip}), and are cited in patents \cite{rouhani2021digitalpatent}. It implies that white-box model watermarking is an active research area. Considering the recent proposal of standards and laws on AI copyright \cite{EU_standards,EU_report}, we believe the need for research on model copyright protection is current and pressing. In this work, we covered nine published white-box watermarking schemes. We leave the evaluation on some more recent watermarking schemes to future work.


\noindent$\bullet$\textbf{ More Threats of Neural Structural Obfuscation.}
During our extensive literature research, we surprisingly discover that not only the white-box watermarking schemes but also other works have strong dependence on the integrity of the neural structure. For example, Chen et al \cite{chen2021copy} present DEEPJUDGE, a testing platform based on non-invasive fingerprint for copyright protection of DNNs, which measures the similarity between the suspect model and victim model using multi-level testing metrics. Despite the authors claiming strong robustness against various attacks in their evaluation, we notice that four out of the six testing metrics in DEEPJUDGE measure the distance between the two models' hidden layer/neuron outputs in the white-box scenario, most of which are also sensitive to the existence of dummy neurons. In our preliminary results reported in Table \ref{tab:deepjudge}, the scheme fails to verify the stolen model under neural structure obfuscation when using the same threshold suggested in DEEPJUDGE \cite{chen2022deepjudge}. The additional results further 
highlight our revealed vulnerability is important for future research works on model copyright verification to circumvent this design pitfall.

\input{tex/tables/deepjudge.tex}




%% file: tex/tables/deepjudge.tex
\begin{table}[t]
    \centering
        \caption[]{Effectiveness of our attack on DEEPJUDGE. The threshold is excerpt from the evaluation result of DEEPJUDGE \cite{chen2022deepjudge}. For each white-box metric, x/y stands for the results w/o. or w/. error handling.}
    \setlength\tabcolsep{2mm}
    \scalebox{0.7}{
    \begin{tabular}{cccccc}
    \toprule          
         \textbf{Metric} & \textbf{w/o. Attack} & \textbf{Threshold} & \textbf{NeuronZero} & \textbf{NeuronClique} & \textbf{NeuronSplit}  \\ \midrule
       NOD &$0$ & $1.79$ & $-/\textbf{3.82}$ & $-/\textbf{259.20}$ & $-/0.03$  \\  
       NAD &$0$ & $6.14$ & $-/\textbf{113.78}$ & $-/\textbf{8.1}$ & $-/1.31$  \\  
       LOD &$0$ & $6.89$ & $-/\textbf{8.16}$ & $-/\textbf{1309.98}$ & $-/1.61$  \\  
       LAD &$0$ & $3.01$ & $-/\textbf{3.17}$ & $-/\textbf{7.97}$ & $-/0.88$  \\  \bottomrule
    \end{tabular}}

    \label{tab:deepjudge}
\end{table}

%% file: tex/cls.tex
\section{Conclusion}
By thoroughly analyzing the protection mechanisms of the state-of-the-art white-box model watermarks, our work for the first time reveals their common and severe security flaw in the resilience against neural structural obfuscation. To validate this, we propose the novel notion of dummy neurons and implement an automatic framework to generate and inject dummy neurons into a given DNN model in a stealthy yet offensive way, which arbitrarily tampers the embedded watermarks while preserving the model utility. Through extensive experiments, we find all nine state-of-the-art white-box model watermarks with claimed robustness against existing removal attacks fail to recognize the original watermark in the protected model after being obfuscated via our neural structural obfuscation attack. As amendments, we discuss possible defenses to strengthen the verification procedures and recover the model from structural obfuscation.   




%% file: tex/ack.tex
\section*{Acknowledgments}
We would like to thank the anonymous reviewers and the shepherd for their insightful comments that helped improve the quality of the paper. This work was supported in part by the National Key Research and Development Program (2021YFB3101200), National Natural Science Foundation of China (61972099, U1736208, U1836210, U1836213, 62172104, 62172105, 61902374, 62102093, 62102091). Min Yang is a faculty of Shanghai Institute of Intelligent Electronics \& Systems, Shanghai Institute for Advanced Communication and Data Science, and Engineering Research Center of Cyber Security Auditing and Monitoring, Ministry of Education, China. Mi Zhang and Min Yang are the corresponding authors.


%% file: tex/appendix.tex
\appendix
\input{tex/app/proof.tex}
\input{tex/app/coverage.tex}

\input{tex/app/eval.tex}

%% file: tex/app/proof.tex
\section{Omitted Proofs}
\label{sec:app:proof}

\begin{proof}[Correctness of \textit{NeuronClique}.] Below, we prove the output of the obfuscated model is the same as the original one after this group of dummy neurons is injected. First, we denote the original outgoing weights and the output of $i$-th neuron in the $l$-th layer are $W^{l}_{i,\text{out}}$ and $h^{l}_i$, respectively. Then the output of the $(l+1)$-th layer after injecting dummy neurons in the $l$-th layer can be written as:
\begin{equation}
h^{l+1}{}' = \text{ReLU}( \sum_{i=1}^{N_l}(W^{l}_{i,\text{out}} \odot h^l_i) + \sum_{k=1}^d (V^{l}_{k,\text{out}} \odot g^l_k)), 
\end{equation}
where $g^l_k = \text{ReLU}(U^{l}_{k,in} \odot h^{l-1})$ is the output of the $k$-th dummy neuron added into the $l$-th layer. As we set the incoming weights for each dummy neuron as identical, we have $g^l_k = g^l_1$ for $k = 1, 2, ..., d$. Combined with other conditions, the contribution of the dummy neurons to the $(l+1)$-th layer is actually equal to $\textbf{0}$, i.e., $\sum _{k=1}^d (V^{l}_{k,out} \odot g^l_k) = \sum _{k=1}^d (V^{l}_{k,out}) \odot g^l_1 = \textbf{0} \odot g^l_1 = \textbf{0}$. As a result, we can further simplify the formulation of $h^{l+1'}$ as follows:
\begin{equation}
h^{l+1'} = \text{ReLU}(\sum _{i=1}^{N_l}(W^{l}_{i,out} \odot h^l_i)+ \textbf{0}) = h^{l+1},
\end{equation}
which indicates that the output of the victim model after we inject a group of dummy neurons generated by NeuronClique is exactly same as before.
\end{proof}

\begin{proof}[Correctness of \textit{NeuronSplit}]
Similar to NeuronClique, we prove that splitting the original neuron into several substitute neurons by NeuronSplit has no unexpected effect on the functionality of the target model. With the selected neuron replacement and dummy neuron injection, the output of the $(l+1)$-th layer can be written as:
$h^{l+1}{}' = \text{ReLU}( \sum _{i=2}^{N_l}(W^{l}_{i,out} \odot h^l_i) + \sum _{k=0}^d (V^{l}_{k,out} \odot g^l_k)), 
$
where $g^l_k = \text{ReLU}(U^{l}_{k,in} \odot h^{l-1}+ b^{l}_k)$ is the output of the $k$-th substitute neuron injected to the $l$-th layer. Because the values of the incoming weights for each substitute neuron are set as identical to the original neuron $n^l_1$, we have $g^l_k = h^l_1$ for $k = 0, 1, ..., d$. Combined with other identities, we formulate the contribution of these substitute neurons to the $(l+1)$-th layer as $\sum _{k=0}^d (V^{l}_{k,out} \odot g^l_k) = \sum _{k=0}^d (V^{l}_{k,out}) \odot h^l_1 = W^l_{1,out} \odot h^l_1$, which is equivalent to $n^l_1$ in the original model. As a result, we can further simplify the formulation of $h^{l+1'}$ as follows:
$
h^{l+1}{}' = \text{ReLU}(\sum _{i=2}^{N_l}(W^{l}_{i,out} \odot h^l_i) + W^l_{1,out} \odot h^l_1)  = h^{l+1},
$
which indicates that the output of the victim model after replacing the selected neuron and injecting a group of dummy neurons generated by NeuronSplit is provably the same as before.
\end{proof}

%% file: tex/app/coverage.tex
\section{Omitted Technical Details}
\subsection{Technical Details of Kernel Expansion}
\label{sec:app:kernel_expansion}
In Section \ref{sec:dn_camouflage}, we obfuscate the kernel shape of the dummy neurons in convolutional layers with vanishing weights for intuition. We provide the technical details of the kernel expansion with non-zero values in this section. Consider a convolutional layer with $N$ neurons, i.e., $\{n_i\}_{i=1}^{N}$, we can split each neuron $n_i$ into two substitute neurons as $n_i'$ and $m_i$, which have the same incoming weight and satisfy the replacement identity, i.e., $W_{i,in}' = U_{i,in} = W_{i,in}$ and $W_{i,out}' + V_{i,out} = W_{i,out}$. As a result, the expanded convolutional layer with $2\times N$ neurons can be denoted as $\{n_i'\}_{i=1}^{N} \cup \{m_i\}_{i=1}^{N}$. Then we can expand the kernel in the outgoing weight of replaced neuron $n_i'$ and $m_i$ with opposite values, while the extra features values in the next layer introduced by the expanded weights of $n_i'$ can be canceled out by the dummy neuron $m_i$, as their outputs in the current layer are exactly the same to each other.

\subsection{On Attack Applicability}
\label{sec:app:coverage}
\noindent\textbf{Dealing with Normalization Layers.}
Similar to fully-connected layers, normalization layers such as batch normalization \cite{ioffe2015batch}, group normalization, and instance normalization implement an elementwise linear transformation on the input $x$: $\hat{x} = \gamma\odot\frac{x - \mu}{\sigma} + \beta,$
where $\gamma$, $\beta$ are learnable parameters and $\mu,\sigma$ are the statistics of the historical training data. Typically, a normalization layer follows a convolutional/fully-connected layer in modern DNN architectures. Therefore, to stay compatible with the dummy neurons injected in the preceding layers, our attack correspondingly expands the normalization layers by assigning the identical coefficients for the inputs from the dummy neurons in the same group.


\noindent\textbf{Dealing with Other Complex Model Architectures.}
Besides, our removal attack can be easily applied to the watermarked DNNs with special connections between the neurons of different layers, e.g. ResNet with shortcuts \cite{he2016resnet} and Inception-V3 with parallel convolution \cite{szegedy2016inception} operations, which have much more complicated architecture than the simple convolutional neural networks. For example, we can obtain the equivalent branches in each Inception block separately to remove the embedded watermark in Inception-V3. More technical details can be found in Appendix \ref{sec:app:coverage}. Only if the adversary knows well about the forwarding computation in the target DNN, which is a common knowledge in white-box watermarking settings, he/she can readily extend our attack with small adjustments, which is left for future works.

In Section \ref{sec:discussion}, we discuss the broad applicability of our attack on other watermarked models with various architecture, i.e., ResNet, not limited to simple convolutional neural network, by carefully setting the injection positions. We provide more detailed analysis and proofs below. First, we combine the $l$-th convolutional layer with possible normalization layer (e.g., batch normalization) as follows:
\begin{equation}
    y^l = \gamma \frac{W_c\odot x -\mu} {\sigma}+ \beta = (\frac {\gamma} {\sigma} \cdot W_c )\cdot x + (\frac {\mu} {\sigma} - \beta),
\end{equation}
as the adversary has full control over the victim model. We denote the weight of the combined convolutional layer as $W' = \frac {\gamma} {\sigma} \cdot W_c $ and $b' = \frac {\mu} {\sigma} - \beta$.

For ResNet \cite{he2016resnet}, we inject the same ratio of dummy neurons into the same position of every layer to confront with the existence of skip connections. As a result, the outputs of the dummy neurons generated by our attacks for a certain layer will produce the same feature maps, as we align the output of dummy neurons from different layers with the same size before the possible shortcut connections.

For Inception-V3 \cite{szegedy2016inception}, the inception modules apply multiple sizes of kernel filters to extract multiple representations, which usually consists of several branches. As a result, we generate the dummy neurons with the output weights which satisfy the cancel-out or replacement identity for each individual branch, and then inject these dummy neurons into each layer as proposed in Section \ref{sec:dn_injection}.

%% file: tex/app/eval.tex
\section{Omitted Evaluation Results}
\label{sec:app:eval}

\noindent\textbf{RIGA.}
Wang et al. \cite{wang2021riga} enhance the covertness and robustness of prior white-box watermarking methods against watermark detection and removal attacks based on adversarial training and more sophisticated transformation function. They train a watermark detector to serve as a discriminator to encourage the distribution of watermark-related weights to be similar to that of unwatermarked models. Meanwhile, they replace the watermark extractor, which has been previously implemented with a predefined linear transformation \cite{uchida2017embedding}, with a learnable fully-connected neural network (FCN), for boosting the encoding capacity of watermarking messages. Similar to Uchida et al.\cite{uchida2017embedding}, the watermark-related weights are first selected from the target model and then projected to a binary string $s'$ via the FCN-based extractor during the ownership verification procedure.

\noindent\textbf{Discussion.}
Simply replacing the linear transformation matrix in Uchida et al. \cite{uchida2017embedding} to a learnable extractor can not completely eliminate the removal threats from our attack based on model structural obfuscation. As a result, RIGA has the similar vulnerability of \cite{uchida2017embedding} as their watermark extraction procedures only differ into the type of extractor, which is also inexecutable due to the incompatible input dimension of the trained extractor for RIGA.

\noindent\textbf{Evaluation Results.}
We follow their evaluation settings to watermark Inception-V3 trained on CelebA, which achieves $95.90\%$ accuracy and $0\%$ BER \cite{code-riga}. We employ the default setups that the watermark is embedded into the third convolutional layer of the target model and the extractor is a multiple layer perceptron with one hidden layer. With our attack framework, we successfully inhibit the ownership verification of RIGA without any loss to the utility of victim model. Even applying the error-handling mechanisms, the BER of extracted message is increased to an unacceptable level. For example, when we utilize Max-First error-handling to obtain the embedded watermark, the BER is increased to $76.04\%$ when we inject the dummy neurons generated via \textit{NeuronSplit}.

\noindent\textbf{Passport-aware Normalization.} 
Zhang et al. \cite{zhang2020passportaware} propose another passport-based watermark method without modifying the target network structure, which would otherwise incur notable performance drops. They adopt a simple but effective strategy by training the passport-free and passport-aware branches in an alternating order and maintaining the statistic values independently for the passport-aware branch at the inference stage. Similar to DeepIPR, the authors design the learnable $\gamma, \beta$ to be relevant to the original model for stronger ownership claim. During the extraction of model watermarks, the transformation function $A$ first projects the $\gamma$ by an additional FCN model to an equal-length vector and then utilizes the signs of the vector to match the target signature.

\noindent\textbf{Discussion.}
While this method improves DeepIPR in terms of model performance by preserving the network structure and improving transformation function $A$ with linear transformation and sign function, we discover it is still inexecutable because of the incompatible dimensions between the extracted watermark and target one.

\noindent\textbf{Evaluation Results.}
When we embed the model watermark into a ResNet18 trained on the CIFAR-100 via passport-aware normalization \cite{code-aware}, we are able to achieve $0\%$ BER, while preventing the original model utility from unacceptable drops. Our proposed structural obfuscation attacks demonstrate sufficient effectiveness to remove this white-box watermark and invalidate the passport-aware branch independently as Fig \ref{fig:scaled_ber} shows. For example, with the error-handling of Max-First, the injection of dummy neurons generated by \textit{NeuronSplit} can boost the BER to $56.17\%$.